\documentclass[11pt]{article}
\usepackage{mathrsfs}
\usepackage{natbib,graphicx,setspace,lscape,longtable}
\usepackage{mathrsfs,amsmath,amsthm,amssymb,color}
\usepackage{natbib,epsfig,graphicx,pdfpages}
\usepackage{epstopdf}
\usepackage{rotating}
\usepackage{subfigure}
\usepackage{graphicx}
\usepackage{url}

\bibpunct{(}{)}{;}{a}{,}{,}

\newtheorem{theorem}{Theorem}
\newtheorem{lemma}{Lemma}
\newtheorem{proposition}{Proposition}

\usepackage{amsthm}

\newtheorem{assumption}{{Assumption}}
\def\beq{\begin{equation}}
\def\eeq{\end{equation}}
\def\beqr{\begin{eqnarray}}
\def\eeqr{\end{eqnarray}}
\def\beqrs{\begin{eqnarray*}}
\def\eeqrs{\end{eqnarray*}}
\def\bet{\begin{theorem}}
\def\eet{\end{theorem}}
\def\bel{\begin{lemma}}
\def\eel{\end{lemma}}
\def\bep{\begin{proposition}}
\def\eep{\end{proposition}}
\def\bg{\begin{figure}[tbph]\begin{center}}
\def\eg{\end{center}\end{figure}}

\def\bc{\begin{center}}
\def\ec{\end{center}}

\def\wt{\widetilde}
\def\wh{\widehat}

\def\bJ{\mathbf J}

\def\diag{\mbox{diag}}

\numberwithin{equation}{section}

\newcommand{\Var}{\textnormal{Var}}
\newcommand{\Cov}{\textnormal{Cov}}

\newcommand{\bA}{{\mathbf A}}

\newcommand{\bF}{{\mathbf F}}

\newcommand{\bG}{{\mathbf G}}
\newcommand{\bH}{{\mathbf H}}
\newcommand{\bI}{{\mathbf I}}

\newcommand{\bL}{{\mathbf L}}
\newcommand{\bM}{{\mathbf M}}
\newcommand{\bQ}{{\mathbf Q}}
\newcommand{\bP}{{\mathbf P}}
\newcommand{\bR}{{\mathbf R}}
\newcommand{\bS}{{\mathbf S}}

\newcommand{\bU}{{\mathbf U}}
\newcommand{\bV}{{\mathbf V}}
\newcommand{\bW}{{\mathbf W}}

\newcommand{\ba}{{\mathbf a}}

\newcommand{\be}{{\mathbf e}}
\newcommand{\bff}{{\mathbf f}}
 \newcommand{\bgg}{{\mathbf g}}
\newcommand{\bh}{{\mathbf h}}

\newcommand{\bq}{{\mathbf q}}
\newcommand{\br}{{\mathbf r}}
\newcommand{\bt}{{\mathbf t}}

\newcommand{\bu}{{\mathbf u}}
\newcommand{\bv}{{\mathbf v}}
\newcommand{\bw}{{\mathbf w}}
\newcommand{\bx}{{\mathbf x}}
\newcommand{\by}{{\mathbf y}}
\newcommand{\bz}{{\mathbf z}}
 
\newcommand{\balpha} {\boldsymbol{\alpha}}

\newcommand{\bfeta}  {\boldsymbol{\eta}}

\newcommand{\bOmega}{\boldsymbol{\Omega}}

\newcommand{\bSigma}{\boldsymbol{\Sigma}}
\newcommand{\bDelta}{\boldsymbol{\Delta}}

\newcommand{\bve}{\mbox{\boldmath$\varepsilon$}}

\newcommand{\bTheta} {\boldsymbol{\Theta}}
\newcommand{\bPhi} {\boldsymbol{\Phi}}

\newcommand{\bPi}{\boldsymbol{\Pi}}

\newcommand{\bxi} {\boldsymbol{\xi}}

\newcommand{\bGamma} {\boldsymbol{\Gamma}}
\newcommand{\bLambda} {\boldsymbol{\Lambda}}

\newcommand{\bD}{{\mathbf D}}

\newcommand{\ve}{{\varepsilon}}
\renewcommand{\epsilon}{{\ve}}
\renewcommand{\hat}{\widehat}

\def\wt{\widetilde}
\newcommand{\nn}{\nonumber}

\newcommand{\beqn}{\begin{eqnarray}}             
\newcommand{\eeqn}{\end{eqnarray}}               
\def\JRSSB{{\sl Journal of the Royal Statistical Society}, {\bf B}}

\def\JASA{{\sl Journal of the American Statistical Association }}

\begin{document}

\title{Modeling High-Dimensional Unit-Root Time Series}

\author{
Zhaoxing Gao, Department of Mathematics, Lehigh University \\  
and Ruey S. Tsay, 
Booth School of Business, University of Chicago
}


\maketitle

\begin{abstract}
This paper proposes a new procedure to build factor models for high-dimensional unit-root time 
series by postulating that a $p$-dimensional unit-root process is a nonsingular linear transformation of a set of unit-root processes, a set of stationary common factors, which are dynamically dependent, and some idiosyncratic white noise components. For the stationary components, 
we assume that the factor process captures the temporal-dependence and the idiosyncratic white noise series explains, jointly with the factors, the cross-sectional dependence. 
The estimation of nonsingular linear loading spaces is carried out in two steps. 
First, we use an eigenanalysis of a nonnegative definite matrix of the data to separate the unit-root processes from the stationary ones and a modified method to specify the number of unit roots. We then employ another eigenanalysis and a projected principal component analysis to identify the stationary common factors and the white noise series. 
We propose a new procedure to specify the number of white noise series and, hence, the 
number of stationary common factors, establish asymptotic properties of the proposed method for both fixed and diverging $p$ as the sample size $n$ increases, and 
use simulation and a real example to demonstrate 
the performance of the proposed method in finite samples. We also compare our method with some commonly used ones in the literature regarding the forecast ability of the extracted factors and find that the proposed method performs well in out-of-sample forecasting of a 508-dimensional PM$_{2.5}$ series in Taiwan.
\end{abstract}


\newpage

\section{Introduction}
High-dimensional data are common in many scientific fields including biology, business, economics, 
and environmental studies.  In many applications, the data consist naturally of  high-dimensional time series and exhibit characteristics of unit-root nonstationarity. For instance, 
the monthly consumer price indexes of European countries tend to exhibit upward trends 
associated with inflation. 
In theory, the vector autoregressive integrated moving-average (VARIMA) models can be used to 
analyze  such a high-dimensional time series, but they often encounter the difficulties of 
high-dimensional co-integration testing, over-parametrization, and lack of identifiability in real applications. See, for instance, \cite{johansen2002} for the first difficulty and 
\cite{TiaoTsay_1989}, \cite{lutkepohl2006}, \cite{Tsay_2014}, and the references therein, for the latter.   Therefore, detecting the number of unit roots and 
dimension reduction become a necessity in analyzing high-dimensional unit-root time series. 
Various methods have been developed in the literature for 
dimension reduction or structural specification of multivariate 
time series analysis, including the scalar component models of \cite{TiaoTsay_1989}, the LASSO regularization in VAR models by \cite{ShojaieMichailidis_2010} and \cite{SongBickel_2011}, the sparse VAR model via partial spectral coherence in \cite{Davis2012}, and the factor modeling by \cite{BaiNg_Econometrica_2002},  \cite{StockWatson_2002a, StockWatson_2005}, \cite{forni2005} and \cite{lamyao2012}, among others. However, most of the 
aforementioned studies focus 
on stationary processes. For unit-root time series, cointegration is often used 
to account for the  common trends and to avoid non-invertibility induced by over-differencing. 
See \cite{engle-granger1987}, \cite{johansen1988, johansen1991}, \cite{Tsay_2014}, and the 
references therein. But the cointegration rank of a multiple time series is unknown in applications, 
and many approaches have been proposed to estimate the unknown rank from data, starting from \cite{engle-granger1987} and the popular likelihood ratio (LR) test in \cite{johansen1988,johansen1991} with a parametric 
integrated VAR setting, to \cite{saikkonen2000} and \cite{aznar2002}. 
As discussed in \cite{johansen2002}, the conventional co-integration tests may fare poorly when the 
dimension of the time seres is high. Yet there exist many applications involving high-dimensional 
time series. For example, \cite{engel-etal2015} contemplated the possibility of determining the cointegration rank of a system of seventeen OECD exchange rates. \cite{banerjee-etal2004} emphasized the importance of testing for no cross-sectional cointegration in panel cointegration analysis, and the cross-sectional dimension of modern macroeconomic panel can easily be as large as several hundreds. Therefore, the complexity of the dynamical dependence in high-dimensional unit-root time series requires further investigation, especially extracting dynamic information from such data plays an important role  in modeling and forecasting large serially dependent data.

This article provides a new approach to analyze high-dimensional unit-root time series from a factor modeling perspective. Like  \cite{zhang-etal2019}, we assume that a $p$-dimensional time series 
is a nonsingular linear transformation of some common unit-root processes and a stationary vector process. However, in contrast to \cite{zhang-etal2019}, but in agreement with \cite{bai2004}, 
we postulate that the number of unit roots is relatively small.   
Due to the nature of a unit-root process, we assume the unit-root factors contribute to both the cross-sectional and temporal dependencies of the data, which are different from those in \cite{bai2004}.
To further reduce the dimensionality, we assume the stationary vector process is a nonsingular linear transformation of certain common stationary factors, which are dynamically dependent, and a vector idiosyncratic white noise series. In other words, for the 
stationary part, we assume the common factors capture all the non-trivial dynamics of the data, but the cross-sectional dependence may be explained by both the common factors and the idiosyncratic components. This is different from the traditional factor analysis where factors capture most of the cross-sectional dependence, while the idiosyncratic terms may contain some non-trivial temporal dependence; see, for example, \cite{BaiNg_Econometrica_2002}. Under the entertained model, 
the common factors explain all the dynamic dependencies of the stationary component of the data and the  idiosyncratic white noises may be contemporaneously correlated with each other 
so that they also contribute to the cross-sectional dependence between the series. 
Therefore, the idiosyncratic white noises of the 
proposed model is also different from the orthogonal factor models in, for example, \cite{maxwell1977} that the idiosyncratic terms (or specific factors therein) tend to be combinations of measurement errors and disturbances that are uniquely associated with the individual variables.   Our approach generalizes that of \cite{gaotsay2019a, gaotsay2019b} by allowing the number of stationary factors to diverge with the sample size. 
To summarize, under the proposed model,  a $p$-dimensional time series 
is a nonsingular linear transformation of certain unit-root  common trends, some stationary common factors which are dynamically dependent, and a white noise idiosyncratic 
process. The proposed model is an extension of the work of \cite{zhang-etal2019} and \cite{gaotsay2019b}, and is in line with the framework of \cite{TiaoTsay_1989} 
because any finite-order $p$-dimensional 
VARMA time series can always 
be written as a nonsingular linear transformation of $p$ scalar component models (SCM) via 
canonical correlation analyses of constructed vector time series. 
See \cite{TiaoTsay_1989} and Section 2.1 below. The only difference is that we 
focus on white noise series, which are SCM of order (0,0), and do not consider specifically individual 
SCM of order beyond (0,0). 
\cite{penaponcela2006} also considered a multiple time series model driven by some common unit-root and some  stationary factors when the dimension is finite. This paper also marks an extension of their approach to high-dimensional nonstationary factor modeling with  certain model structure.

Although the maximum likelihood method is more efficient than 
the principal components method in applying the traditional factor model, it is not feasible for the  
proposed model because our factors not only explain the variance of the data, but also capture the dynamic dependencies, and the covariance matrix of the idiosyncratic term may not be 
diagonal as in  \cite{baili2012}. 
Instead, similar to \cite{po1988}, \cite{robinson2002}, \cite{penaponcela2006}, and \cite{zhang-etal2019}, we employ methods based on eigenanalysis. We first estimate the number of unit-root factors (or equivalently the cointegration rank) and extract them from the data by an  eigenanalysis of a nonnegative definite matrix, which is a function of the sample covariance and lagged autocovariance matrices of the data. The nonnegative definite matrix used in this paper is different from the sample covariance used in \cite{bai2004} and the fixed lag sample covariance or autocovariances used in \cite{penaponcela2006}, because we adopt a combination of the covariance and some lagged autocovariances together to capture simultaneously the cross-sectional and temporal dependence of the data. In addition, we propose to use an average of the absolute autocorrelations of the transformed components to identify the number of unit-root series. The absolute value can avoid the impact of sign changes in the autocorrelations introduced  by the stationary part embedded in the individual unit-root component. Limited simulation studies suggest that, although the performance of the proposed method is comparable with the one in \cite{zhang-etal2019} when the sample size is sufficiently large, the use of absolute autocorrelations can improve the accuracy in estimating the number of unit-root processes, especially when the sample size is small. An accurate specification of the number of unit-root factors can provide more accurate information for the second eigenanalysis, from which the number of stationary 
common factors is identified. Specifically, to estimate the number of stationary common factors in 
the second eigenanalysis, we apply the method of \cite{gaotsay2019b} to the transformed data which are orthogonal to the unit-root components. Since eigenanalysis 
is mainly based on spectral decomposition of a nonnegative definite matrix, the resulting  
components are arranged according to the amount of variabilities explained. The ordering 
of the components thus provides no information concerning the temporal dependence in each 
principal component. Consequently, to specify the number of white noises and, hence, the number of 
stationary common factors,  we propose to reorder the transformed components of 
eigenanalysis according to their $p$-values (in ascending order) of the Ljung-Box statistic in 
testing their serial correlations.  Limited experience shows that this re-ordering 
procedure is helpful in detecting the number of white noise series when the dimension is large 
and the sample size is small. 

Under the proposed framework, the dimension of the idiosyncratic white noise may go to infinity 
and some largest eigenvalues of the covariance matrix of the white noise may also diverge. 
We refer to the latter case as prominent noise effect and apply the projected 
principal component analysis of \cite{gaotsay2019b} to mitigate the effect of such prominent noises  
in estimating the stationary common factors. Consequently, our proposed method 
can successfully separate  the nonstationary unit-root processes, the stationary common factors, 
and the idiosyncratic white noise components.  
In estimating the number of unit-root series (or equivalently the cointegration rank), we could allow the dimension $p$ to grow as fast as the sample size $n$. This relaxes the constraint of 
\cite{zhang-etal2019} and the error-correction factor models of \cite{tu-etal2019} that $p$ can 
grow  at most with $\sqrt{n}$. Asymptotic properties of the proposed  method are established for both fixed $p$ and diverging $p$ as the sample size $n$ tends to infinity.

The rest of the paper is organized as follows. We introduce the proposed model and 
estimation methodology in Section 2.  
 In Section 3, we study the theoretical properties of the proposed model and its associated 
 estimates. 
Section 4 illustrates the performance of the proposed model using both 
simulated and real data sets, including analysis of a 508-dimensional time series of 
PM$_{2.5}$ measurements in Taiwan with 744 observations. 
Section 5 provides some discussions and concluding remarks. 
All technical proofs are relegated to an Appendix. 
Throughout the article,
 we use the following notation: $||\bu||_2 = (\sum_{i=1}^{p} u_i^2)^{1/2} $
is the Euclidean norm of a $p$-dimensional vector  
$\bu=(u_1,..., u_p)'$, $\|\bu\|_\infty=\max_i |u_i|$, and $\bI_k$ denotes the $k\times k$ identity matrix. For a matrix $\bH=[h_{ij}]$,  $\|\bH
\|_2=\sqrt{\lambda_{\max} (\bH' \bH ) }$ is the operator norm, where
$\lambda_{\max} (\cdot) $ denotes the largest eigenvalue of a matrix, and $\|\bH\|_{\min}$ is the square root of the minimum non-zero eigenvalue of $\bH'\bH$. The superscript $'$ denotes 
the transpose of a vector or matrix. 
Finally, we use the notation $a\asymp b$ to denote $a=O(b)$ and $b=O(a)$.

\section{The Proposed Methodology}
\subsection{The Setting and Framework}
Let $\by_t=(y_{1t},...,y_{pt})'$ be a $p$-dimensional $I(1)$ time series process. 
We assume $\by_t$ is observable and admits the following latent structure
\begin{equation}\label{int:eq}
\by_t=\bL\left[\begin{array}{c}
\bff_{1t}\\
\bff_{2t}\\
\bve_t
\end{array}\right]=[\bL_1,\bL_2,\bL_3]\left[\begin{array}{c}
\bff_{1t}\\
\bff_{2t}\\
\bve_t
\end{array}\right]=\bL_1\bff_{1t}+\bL_2\bff_{2t}+\bL_3\bve_t,
\end{equation}
where $\bL\in R^{p\times p}$ is a full rank loading matrix, $\bff_{1t}=(f_{1,1t},\ldots,f_{1,r_1t})'$ 
is an $r_1$-dimensional $I(1)$ process, $\bff_{2t}=(f_{2,1t},\ldots,f_{2,r_2t})'$ is an $r_2$-dimensional stationary process, and $\bve_t=(\epsilon_{1t},\ldots,\epsilon_{vt})$ is a $v$-dimensional white noise series with $v=p-r$, $r=r_1+r_2$, and $r_i \geq 0$. For meaningful dimension reduction, we assume $r_1$ is a relatively small and fixed integer as that in \cite{bai2004}, and  $r_2$ can be either fixed or  slowly growing  with the dimension $p$, which extends the results in \cite{gaotsay2019b}. 
In addition, we also assume that $\bff_{2t}$ and $\bve_t$ are
independent of each other with $\Cov(\bff_{2t})=\bI_{r_2}$  and $\Cov(\bve_t)=\bI_v$, 
and no linear combination of $\bff_{1t}$ is a stationary process and no linear combination 
of $\bff_{2t}$ is a white noise. In theory, \Cov($\bff_{1t}$) is time-varying 
because $\bff_{1t}$ consists of unit-root processes, but its sample version may assume 
an identity matrix when the sample size is given and the processes are assumed to 
start at $t$ = 0 with fixed starting values.

The decomposition of (\ref{int:eq}) is general and in line with the framework 
of \cite{TiaoTsay_1989}. 
Under the scalar component models of \cite{TiaoTsay_1989}, any finite-order 
$p$-dimensional VARIMA($p,d,q$) process consists of $p$ scalar component models (SCM) 
of orders $(p_i,q_i)$ with $p_i+d_i \leq p+d$ and 
$q_i \leq q$. The SCM and the associated transformation matrix can be obtained via 
canonical correlation analyses between constructed random vectors of $\by_t$ 
and its lagged variables. Model (\ref{int:eq}) can be considered as consisting of 
three classes of SCM, namely, the unit-root processes, the processes of SCM(0,0), and 
the process of stationary SCM($p_i,q_i$) with $p_i+q_i > 0$. Specifically, 
let $\bV'=\bL^{-1}$ and $\bV=(\bV_1,\bV_2,\bV_3)$ with $\bV_1\in R^{p\times r_1}$, $\bV_2\in R^{p\times r_2}$ and $\bV_3\in R^{p\times v}$. Model (\ref{int:eq}) is a transformation model employed in  \cite{TiaoTsay_1989}. 

Another way to see the generality of Model (\ref{int:eq}) is to employ the canonical correlation 
analysis of \cite{BoxTiao_1977} involving a VARIMA process $\by_t$ and its 
1-step ahead prediction at time index $t-1$. Under such an analysis, $(\bff_{1t}', \bff_{2t}', \bve_t')'$ are the canonical variates whose correlations with the past values are in descending order, and the correlation that is close to unity corresponds to a unit-root nonstationary series, and those close to zero are close to a white noise; see also \cite{penaponcela2006}. As a result, we consider the top canonical variates $\bff_{1t}$ a unit-root process, the bottom $\bve_t$ a white noise, and the ones in the middle are dynamically dependent stationary processes. Consequently, the columns of $(\bV_2,\bV_3)$ are the cointegrating vectors of $\by_t$. 
For more details, we refer interested readers to \cite{BoxTiao_1977} and \cite{TiaoTsay_1989} for general discussions. 

To illustrate the identification issue of Model (\ref{int:eq}) and to provide a concrete analysis, 
we let $\bgg_{2t}=(\bff_{2t}',\bve_t')'$ and $\bG_2=[\bL_2,\bL_3]$, and rewrite 
Model (\ref{int:eq}) as
\begin{equation}\label{2:eq}
\by_t=[\bL_1,\bG_2]\left[\begin{array}{c}
\bff_{1t}\\
\bgg_{2t}
\end{array}\right],
\end{equation}
where $\bgg_{2t}$ is a $(p-r_1)$-dimensional stationary process.
Note that Model (\ref{2:eq}) is not uniquely defined, as $[\bL_1,\bG_2]$ and $(\bff_{1t}',\bgg_{2t}')'$ can be replaced by $[\bL_1,\bG_2]\bH^{-1}$ and $\bH(\bff_{1t}',\bgg_{2t}')'$, respectively, for any invertible $\bH$ with the form
\begin{equation}\label{H}
\bH=\left[\begin{array}{cc}
\bH_{11}&\bH_{12}\\
\bf 0&\bH_{22}
\end{array}\right],
\end{equation}
where $\bH_{11}$ and $\bH_{22}$ are square matrices of sizes $(p-r_1)$ and $r_1$, respectively. In other words, the nonstationary components can include any linear combinations of the stationary ones. However, for any nonorthogonal invertible matrix $[\bL_1,\bG_2]$, we always have the decomposition $[\bL_1,\bG_2]=\bA\bU$, where $\bA$ is orthonormal and $\bU$ is upper-triangular, and we may replace  $[\bL_1,\bG_2]$ and $(\bff_{1t}',\bgg_{2t}')'$ by $\bA$ and $\bU(\bff_{1t}',\bgg_{2t}')'$ without altering the structure of the model. Let $\bA=[\bA_1,\bA_2]$ and $(\bx_{1t}',\bx_{2t}')'=\bU(\bff_{1t}',\bgg_{2t}')'$. There is no loss of generality in assuming that
\begin{equation}\label{w:eq}
\by_t=\bA\left[\begin{array}{c}
\bx_{1t}\\
\bx_{2t}
\end{array}\right]=[\bA_1,\bA_2]\left[\begin{array}{c}
\bx_{1t}\\
\bx_{2t}
\end{array}\right],
\end{equation}
where $\bA$ is an orthonormal matrix, $\bx_{1t}$ is an $r_1$-dimensional $I(1)$ process, and $\bx_{2t}$ is a $(p-r_1)$-dimensional stationary process. Therefore, $\bx_{1t}=\bA_1'\by_t$ and $\bx_{2t}=\bA_2'\by_t$. For any $\bH$ in the form of (\ref{H}) to be orthonormal, we can show that $\bH$ is a block-orthonormal matrix. Thus,  Model (\ref{w:eq}) is still not identifiable and $\bA_1$ and $\bA_2$ cannot be uniquely defined. However, the linear spaces spanned by the columns of $\bA_1$ and $\bA_2$, denoted by $\mathcal{M}(\bA_1)$ and $\mathcal{M}({\bA_2})$, can be uniquely defined.

To proceed with the proposed dimension reduction procedure, 
noting that $(\bx_{1t}',\bx_{2t}')'=\bU(\bff_{1t}',\bgg_{2t}')'$ symbolically for an upper triangular matrix $\bU$, we further assume that 
\begin{equation}\label{st:eq}
\bx_{2t}=\bU_{22}\left[\begin{array}{c}
\bff_{2t}\\
\bve_t
\end{array}\right]=\bU_{22,1}\bff_{2t}+\bU_{22,2}\bve_t,
\end{equation}
where $\bU_{22}=[\bU_{22,1},\bU_{22,2}]$ is the lower diagonal block of $\bU$ in the form of (\ref{H}). Given Models (\ref{w:eq}) and (\ref{st:eq}), we estimate $r_1$, $r_2$, the linear spaces $\mathcal{M}(\bA_1)$, $\mathcal{M}(\bA_2)$, and $\mathcal{M}(\bA_2\bU_{22,1})$ as well as 
recover the processes $\bx_{1t}$ and $\bff_{2t}$.

\subsection{Estimation Methods}
To introduce estimation, we  assume first that $r_1$ and $r_2$ are known and estimate $\bA_1$, $\bA_2$, and $\bU_{22,1}$, or equivalently the linear spaces spanned by their columns. The estimation 
of $r_1$ and $r_2$ is given in Section 2.3 below. We start with the case that $p$ is finite. 
For $k\geq 0$, we define the lag-$k$ sample covariance matrix of $\by_t$ as 
\begin{equation}\label{sigy}
\wh\bSigma_y(k)=\frac{1}{n}\sum_{t=k+1}^n(\by_{t}-\bar{\by})(\by_{t-k}-\bar{\by})',\quad\bar{\by}=\frac{1}{n}\sum_{t=1}^n\by_t.
\end{equation}
For any $\ba_1\in\mathcal{M}(\bA_1)$ and $\ba_2\in\mathcal{M}(\bA_2)$, $\ba_1'\wh\bSigma_y(k)\ba_1$ is the lag-$k$ sample autocovariance  
of the $I(1)$ process $\ba_1'\by_t$, and $\ba_2'\wh\bSigma_y(k)\ba_2$ is that of the weakly stationary univariate time series $\ba_2'\by_t$. When $p$ is finite, it is not hard to see that $\ba_2'\wh\bSigma_y(k)\ba_2$ converges to a finite constant almost surely under some mild conditions, and with probability tending to one that
\begin{equation}\label{auto:order}
0<\ba_1'\wh\bSigma_y(k)\ba_1\leq \left\{\begin{array}{ll} Cn & \mbox{if $E(\ba_1'\by_t) = 0$}, \\ Cn^2 & \mbox{if $E(\ba_1'\by_t) \neq 0$,}\end{array}\right.
\end{equation}
for some constant $0< C<\infty$. See Theorems 1 and 2 of \cite{penaponcela2006}. 
Hence, the $r_1$ directions in the space $\mathcal{M}(\bA_1)$ make $\ba_1'\wh\bSigma_y(k)\ba_1$ as large as possible for all $k\geq 0$. 

\subsubsection{Estimation of Unit-Root Processes}
Similar to several works in the literature, e.g., \cite{lamyao2012}, we combine the information over different lags of $\by_t$ and define
\begin{equation}\label{wy}
\wh\bM_1=\sum_{k=0}^{k_0}\wh\bSigma_y(k)\wh\bSigma_y(k)',
\end{equation}
where $k_0\geq 1$ is a prespecified fixed integer. We use the product $\wh\bSigma_y(k)\wh\bSigma_y(k)'$ instead of $\wh\bSigma_y(k)$ to ensure that each term in the sum of (\ref{wy}) is nonnegative definite, and there is no information cancellation over different lags. Unlike \cite{bai2004} and \cite{penaponcela2006}, $\wh\bM_1$ of (\ref{wy}) combines the sample covariance and lag autocovariances together to focus on both the cross-sectional and the dynamic dependence of the data simultaneously. Limited experience suggests that a relatively small $k_0$ is sufficient in providing useful information concerning the model structure of $\by_t$, because, for the stationary component $\bx_{2t}$, cross-correlation matrices decay to zero exponentially as $k$ increases. Also, the choice of $k_0$ seems to be not sensitive. See, for instance, the simulation results in Section 4. It can be shown that the $r_1$ largest eigenvalues of $\wh\bM_1$ are at least of order $n^2$, while the other $(p-r_1)$ eigenvalues are $O_p(1)$. Hence, $\mathcal{M}(\bA_1)$ can be estimated by the linear space spanned by the $r_1$ eigenvectors of $\wh\bM_1$ corresponding to the $r_1$ largest eigenvalues, and $\mathcal{M}(\bA_2)$ can be estimated by that spanned by the $(p-r_1)$ eigenvectors of $\wh\bM_1$ corresponding to the $(p-r_1)$ smallest eigenvalues.

Let $(\wh\ba_{1,1},...,\wh\ba_{1,r_1},\wh\ba_{2,1},...,\wh\ba_{2,p-r_1})$ be the orthonormal eigenvectors of $\wh\bM_1$ corresponding to the eigenvalues arranged in descending order. Define $\wh\bA_1=(\wh\ba_{1,1},...,\wh\ba_{1,r_1})$ and $\wh\bA_2=(\wh\ba_{2,1},...,\wh\ba_{2,p-r_1})$, the estimated $\wh\bx_{1t}$ and $\wh\bx_{2t}$ are given by 
\begin{equation}\label{xhat}
\wh\bx_{1t}=\wh\bA_1'\by_t\quad\text{and}\quad\wh\bx_{2t}=\wh\bA_2'\by_t.
\end{equation}
Then, $\mathcal{M}(\widehat{\bA}_1)$ and $\mathcal{M}(\wh{\bA}_2)$, the linear spaces spanned by the eigenvectors of $\wh\bM_1$, are consistent estimators for $\mathcal{M}(\bA_1)$ and $\mathcal{M}(\bA_2)$, respectively. See Theorem 1 below for details.

When $p$ is diverging, it is reasonable to consider strengths of the factors $\bx_{1t}$ and $\bx_{2t}$, and the strengths from the columns of $\bL_1$ and $\bG_2$. See the discussion in \cite{gaotsay2019b} for details. For simplicity, we introduce a parameter $\delta\in[0,1)$ such that the nonzero singular values of $\bL_1$ and $\bL_2$, and a few largest singular values 
of $\bL_3$ are of order $p^{(1-\delta)/2}$. It is not hard to see that the order of $\ba_2'\wh\bSigma_y(k)\ba_2$ is $O_p(p^{1-\delta})$ under some mild conditions, and with probability tending to one that $0<\ba_1'\wh\bSigma_y(k)\ba_1\leq Cp^{1-\delta}n$ or $Cp^{1-\delta}n^2$ for some $0<C<\infty$, depending on whether $E(\ba_1'\by_t)=0$ or not. Therefore, $\mathcal{M}(\bA_1)$ can still be estimated by the linear space spanned by the $r_1$ eigenvectors of $\wh\bM_1$ corresponding to the $r_1$ largest eigenvalues, and $\mathcal{M}(\bA_2)$ can be estimated by that spanned by the remaining $(p-r_1)$ eigenvectors of $\wh\bM_1$ corresponding to the  $(p-r_1)$ smallest eigenvalues. The estimators for $\bx_{1t}$ and $\bx_{2t}$ are the same as those in (\ref{xhat}).

\subsubsection{Estimation of Stationary Common Factors}
Turn to the estimation of $\bU_{22,1}$ and $\bff_{2t}$. Because $\bU_{22,1}$ is not uniquely defined and we can replace $(\bA_{2},\bU_{22,1})$ by $(\bA_{2}\bH_1',\bH_1\bU_{22,1})$ or replace $(\bU_{22,1},\bff_{2t})$ by $(\bU_{22,1}\bH_2',\bH_2\bff_{2t})$ for any orthonormal matrices $\bH_1\in R^{(p-r_1)\times (p-r_1)}$ and $\bH_2\in R^{r_2\times r_2}$ without altering the relations. Therefore, only $\mathcal{M}(\bA_{2}\bU_{22,1})$ can be estimated.
We decompose $\bU_{22,1}$ and $\bU_{22,2}$ as $\bU_{22,1}=\bU_1\bQ_1$ and $\bU_{22,2}=\bU_2\bQ_2$, where $\bU_{1}$ and $\bU_2$ are two half orthonormal matrices in the sense that $\bU_1'\bU_1=\bI_{r_2}$ and $\bU_2'\bU_2=\bI_{p-v}$. Then, Model (\ref{st:eq}) can be written as
\begin{equation}\label{re:st}
\bx_{2t}=\bU_1\bz_{2t}+\bU_2\be_t,
\end{equation} 
where $\bz_{2t}=\bQ_1\bff_{2t}$ and $\be_t=\bQ_2\bve_t$. Equivalently, we estimate 
$\mathcal{M}(\bA_2\bU_1)$ and recover $\bz_{2t}$.
First, we can apply the method of \cite{gaotsay2019b} to estimate the linear space $\mathcal{M}(\bU_1)$ and recover $\bz_{2t}$. Specifically, letting $\wh\bx_{2t}=\wh\bA_2'\by_t$, we  
estimate $\mathcal{M}(\bU_1)$ using the transformed data $\wh\bx_{2t}$ because 
$\mathcal{M}(\wh\bA_2)$ is a consistent estimator for $\mathcal{M}(\bA_2)$. Define
\begin{equation}\label{M2}
\wh\bM_2=\sum_{j=1}^{j_0}\wt\bSigma_2(j)\wt\bSigma_2(j)', 
\end{equation}
where $\wt\bSigma_2(j)$ is the lag-$j$ sample autocovariance matrix of $\wh\bx_{2t}$ 
and $j_0$ is a prespecified and fixed positive integer. Assume that $\bV_1$ and $\bV_2$ are the orthogonal complements of $\bU_1$ and $\bU_2$, respectively. From (\ref{re:st}), we see that $\bV_1'\bx_{2t}=\bV_1'\bU_2\be_t$ and $\bV_2'\bx_{2t}=\bV_2'\bU_1\bz_{2t}$ are uncorrelated with each other, and therefore, $\bV_2$ consists of the eigenvectors corresponding to the $r_2$ zero eigenvalues of $\bS:=\bSigma_2\bV_1\bV_1'\bSigma_2$, where $\bSigma_2=\Var(\bx_{2t})$. This leads to the following projected PCA method. Let the columns of $[\wh\bU_1,\wh\bV_1]$ be the orthonormal eigenvectors of $\wh\bM_2$ corresponding to the eigenvalues arranged in descending order, where $\wh\bU_{1}$ contains the first $r_2$ columns and $\wh\bV_{1}$ contains the 
remaining $(p-v)$ columns, and  $\wt\bSigma_2$ be the sample covariance of $\wh\bx_{2t}$. Then the estimation of $\bV_2$ is based on the eigenanalysis of 
\begin{equation}\label{Spca}
\wh\bS=\wt\bSigma_2\wh\bV_1\wh\bV_1'\wt\bSigma_2,
\end{equation}
which is a projected PCA and $\wh\bS$ is an estimator of $\bS$. That is, we project the data $\wh\bx_{2t}$ onto the direction of $\wh\bV_1$, then perform the PCA between $\wh\bx_{2t}$ and the projected coordinates. 
When $p$ is finite, assume $\wh\bV_2$ contains the last $r_2$ eigenvectors corresponding the the smallest $r_2$ eigenvalues of $\wh\bS$, then from the relation $\bV_2'\bx_{2t}=\bV_2'\bU_1\bz_{2t}$, the recovered $\bz_{2t}$ process is
\begin{equation}\label{z2hat}
\wh\bz_{2t}=(\wh\bV_2'\wh\bU_1)^{-1}\wh\bV_2'\wh\bx_{2t}.
\end{equation}

When $p$ is large, similar as that in \cite{gaotsay2019b}, we assume the largest $K$ eigenvalues of the covariance of $\bU_2\be_t$ are diverging, and therefore, the largest $K$ 
eigenvalues of $\wh\bS$ are also diverging. Suppose $\wh\bV_2^*$ contains $(p-r_1-K)$ eigenvectors of $\wh\bS$ corresponding to its $(p-r_1-K)$ smallest eigenvalues, we choose $\wh\bV_2$ as $\wh\bV_2=\wh\bV_2^*\wh\bR$, where $\wh\bR=[\wh\br_1,...,\wh\br_{r_2}]\in R^{(p-r_1-K)\times r_2}$ with $\wh\br_i$ being the eigenvector associated with the $i$-th largest eigenvalues of $\wh\bV_2^*{'}\wh\bU_1\wh\bU_1'\wh\bV_{2}^*$. This choice of estimator guarantees that the matrix $(\wh\bV_2'\wh\bU_1)^{-1}$ behaves well in recovering the factor $\wh\bz_{2t}$. The resulting estimator $\wh\bz_{2t}$ is the same as that in (\ref{z2hat}).  

\subsubsection{Prediction}
With $\wh\bA_1$, $\wh\bA_2$, $\wh\bU_1$, and the estimated nonstationary factor process $\wh\bx_{1t}$ and the stationary one $\wh\bz_{2t}$, we compute the $h$-step ahead prediction of the $\by_t$ series using the formula $\wh\by_{n+h}=\wh\bA_1\wh\bx_{1,n+h}+\wh\bA_2\wh\bU_1\wh\bz_{2,n+h}$, where $\wh\bx_{1,n+h}$ and $\wh\bz_{2,n+h}$ are $h$-step ahead forecasts for $\bx_{1t}$ and $\bz_{2t}$ based on the estimated past values $\{\wh\bx_{11},...,\wh\bx_{1n}\}$ and $\{\wh\bz_{21},...,\wh\bz_{2n}\}$, respectively. This can be done, for example, by fitting a VAR model to $\{\wh\bx_{11},...,\wh\bx_{1n}\}$ and $\{\wh\bz_{21},...,\wh\bz_{2n}\}$, respectively. In addition, we may also adopt the factor-augmented error correction model in \cite{banerjee-etal2014} to do forecasting using the nonstationary factors $\wh\bx_{1t}$ if we are interested in some particular components of $\by_t$.

\subsection{Determination of $r_1$ and $r_2$}
Turn to the estimation of $r_1$ and $r_2$, which are unknown in practice. 
We begin with the estimation of the number of unit roots $r_1$. 
Note that the components of $\wh\bx_t=\wh\bA'\by_t=(\wh x_{1t},\ldots,\wh x_{pt})'$, defined in (\ref{xhat}), are arranged according to the descending order of  eigenvalues of $\wh\bM_1$. Therefore, the order of the components reflects inversely the closeness to stationarity of the component series, with $\{\wh x_{pt}\}$ being most likely stationary, and $\{\wh x_{1t}\}$ being most likely an $I(1)$ process. Based on this observation, we consider a modified 
method to estimate the number of nonstationary components $r_1$. 
\cite{zhang-etal2019} use the average of sample autocorrelations of each transformed 
component to determine its stationarity. Our proposed method makes two important modifications, 
because (a) limited simulation studies show that some transformed stationary components also have large autocorrelations when the dimension is high, especially at the lower-order lags, 
and (b) the stationary part included in a unit-root process may  change the sign 
of its autocorrelations. Our modified method is given below. 
 Let $k_1=1$ and $k_j=k_1+(j-1)l$ for some constant $l\geq 1$. Define
\begin{equation}\label{sm}
S_i(l,m)=\sum_{j=1}^m|\wh\rho_i(k_j)|,
\end{equation}
where $\wh\rho_i(j)$ is the lag-$j$ sample autocorrelation function (ACF) of $\wh x_{it}$. 
 If $\wh x_{it}$
 is stationary, then under some mild conditions, $\hat{\rho}_i(k)$ decays exponentially as $k$ increases. Therefore, $\lim_{m\rightarrow\infty} S_i(l,m)<\infty$ in probability. On the other hand, 
 if  $\wh x_{it}$ is unit-root nonstationary, $\wh\rho_i(k)\rightarrow 1$ in probability for any fixed $k$ as $n\rightarrow\infty$. Therefore, $\lim_{m\rightarrow\infty} S_i(l,m)=\infty$ as $n\rightarrow 
 \infty$. 
 
We use a gap of size $l\geq 1$ in (\ref{sm}) to reduce the effect of high-dimensionality on the 
lower-order sample ACFs of  transformed stationary components, and employ the absolute ACFs 
to avoid the effect of sign changes in ACFs due to the existence of stationary part embedded in the unit-root components. Using the statistic $S_i(l,m)$, we propose the following thresholding 
procedure to estimate $r_1$.  
We start with $i=1$, if the average of absolute sample autocorrelations 
$S_i(l,m)/m\geq c_0$ for some constant $0<c_0<1$, the $\hat{x}_{it}$ series has a unit root 
and increase $i$ by 1 to repeat the detecting process. This detecting process is continued 
until $S_i(l,m)/m< c_0$ or $i=p$. If $S_i(l,m) \geq c_0$ for all $i$, then $\hat{r}_1=p$; otherwise, 
an estimator of $r_1$ is $\wh r_1=i-1$. In our numerical experiments of Section 4, 
we set $c_0=0.3$, $l=3$ and $m=10$, and the estimator $\wh r_1$ performs very well. Consequently, $\wh\bA_1$ and $\wh\bA_2$ in (\ref{xhat}) can be replaced by $\wh\bA_1=(\wh\ba_{1,1},...,\wh\ba_{1,\wh r_1})$ and $\wh\bA_2=(\wh\ba_{2,1},...,\wh\ba_{2,p-\wh r_1})$.
 
 
Next, turn to the estimation of $r_2$, which is the number of stationary common factors. 
Because Model (\ref{re:st}) is the same as (2.2) in \cite{gaotsay2019b} and $\mathcal{M}(\wh\bA_2)$ is a consistent estimator for $\mathcal{M}(\bA_2)$, we apply the white noise test procedure there to the transformed data $\wh\bx_{2t}$ to estimate $v$ and use 
$\wh r_2 = p-\wh r_1 - \wh v$. If the dimension $p$ is small (say less than 10), 
we use a bottom-up procedure to determine the number of 
white noise components. Specifically, we test the null hypothesis that $\wh\xi_{it}$ is a white noise starting with $i=p-\wh r_1$ using, for example, the well-known Ljung-Box statistic $Q(m)$. Clearly, this testing process can only last until $i=1$. If all transformed series $\wh \xi_{it}$ are white noise, then $\wh r_2=0$ and $\wh v=p-\wh r_1$. In general, if $\wh \xi_{it}$ is not a white noise series but $\wh \xi_{jt}$ are for $j=i+1,...,p$, then $\wh r_2=i$ and $\wh v=p-\wh r_1-i$, and we have $\wh\bW=[\wh\bU_1,\wh\bV_1]$, where $\wh\bU_1\in R^{(p-\wh r_1)\times \wh r_2}$ and $\wh\bV_1\in R^{(p-\wh r_1)\times \wh v}$.

For large $p$, we propose a modification before 
conducting the high-dimensional white noise test. 
Specifically, let $\wh \bW$ be the matrix of eigenvectors (in the decreasing order of eigenvalues) of the sample matrix $\wh \bM_2$ in Equation (\ref{M2}) and $\wh\bxi_t=\wh\bW'\wh\bA_2'\by_t=(\wh\xi_{1t},...,\wh\xi_{p-\wh r_1,t})'$ be the transformed series. 
Note that the ordering of components in $\wh\bxi_t$ is based on the eigenvalues which do not have 
any specific association with the temporal dependence of its components. Yet our goal is 
to detect the number of white noise components. It is then reasonable to re-order the 
components of $\wh\bxi_t$ based on the $p$-values of the Ljung-Box statistics of testing 
zero serial correlations in the components of $\wh\xi_t$. 
This re-ordering step enables us to conduct the 
high-dimensional white noise test more efficiently starting with all $p-\wh r_1$ components. 
For simplicity, let $\wh\bxi^*_t$ be the re-ordered series of $\wh\bxi_t$, which is of 
dimension $p-\wh r_1$, and $\wh \bW^*$ be the corresponding matrix of eigenvectors. 
Our high-dimensional white-noise test starts with 
the null hypothesis that $\wh\bxi^*_t$ is  
white noise. If the null hypothesis is rejected, we remove the first component from $\wh\bxi^*_t$ 
and repeat the testing procedure. If all null hypotheses cannot be rejected, then the number of 
white noises is $\wh v = p-\wh r_1$. In general, $\wh v$ is the dimension of the 
subset of $\wh\bxi^*_t$ for which the white-noise hypothesis cannot be rejected. 
The number of stationary common factors is then $\wh r_2 = p-\wh r_1-\wh v$. 
Once $\wh v$ and $\wh r_2$  are estimated, $\wh \bU_1$ and $\wh\bV_1$ can be estimated 
accordingly using columns of $\wh\bW^*$. Simulation studies in Section 4.2 suggest that the performance of detecting the number of stationary factors in both small and large dimensions benefit from this reordering procedure, especially when the sample size is small.

\section{Theoretical Properties}
 In this section, we investigate some asymptotic theory for the estimation methods used in the paper. Starting with the assumption that the number of common factors $r_1$ and $r_2$ are known, we divide the derivations into two cases depending on the dimension $p$. The case of estimated $r_1$ and $r_2$ is discussed later. In the derivation, we adopt the discrepancy measure used by
\cite{panyao2008}: for two $p\times r$ half orthogonal
matrices ${\bf H}_1$ and ${\bf H}_2$ satisfying the condition ${\bf
H}_1'{\bf H}_1={\bf H}_2'{\bf H}_2=\bI_{r}$, the difference
between the two linear spaces $\mathcal{M}({\bf H}_1)$ and
$\mathcal{M}({\bf H}_2)$ is measured by
\begin{equation}
D({\bf H}_1,{\bf
H}_2)=\sqrt{1-\frac{1}{r}\textrm{tr}({\bf H}_1{\bf H}_1'{\bf
H}_2{\bf H}_2')}.\label{eq:D}
\end{equation}
Note that $D({\bf H}_1,{\bf H}_2) \in [0,1].$ 
It is equal to $0$ if and only if
$\mathcal{M}({\bf H}_1)=\mathcal{M}({\bf H}_2)$, and to $1$ if and
only if $\mathcal{M}({\bf H}_1)\perp \mathcal{M}({\bf H}_2)$. Denote the $I(1)$ factors  by $\bx_{1t}=(x_{1,1t},\ldots,x_{1,r_1t})'$ and define $\bw_t=(w_{1t}, \ldots, w_{r_1t})'$, where  $w_{it}\equiv \nabla x_{1,it}$, which is $I(0)$ for $1\leq i\leq r_1$.
For simplicity, we assume $\mathrm{E}\bw_t=\bf 0$ and denote the vector of partial sums 
of the components of $\bw_t$ by
$\bS_n^{r_1}(\bt)\equiv (S_n^1(t_1),...,S_n^{r_1}(t_{r_1}))'=\left(\frac{1}{\sqrt{n}}\sum_{s=1}^{[nt_1]}w_{1s},...,\frac{1}{\sqrt{n}}\sum_{s=1}^{[nt_{r_1}]}w_{r_1s}\right)',$
where $0< t_1<...<t_p\leq 1$ are constants and $\bt=(t_1,...,t_{r_1})'$. 
We always assume that the processes $\bw_t$ and $(\bx_{2t},\bff_{2t})$ are weakly stationary 
and $\alpha$-mixing, that is, their 
mixing coefficients $\alpha_p(k)\rightarrow 0$ as $k\rightarrow\infty$, where
$\alpha_p(k)=\sup_{i}\sup_{A\in\mathcal{F}_{-\infty}^i,B\in \mathcal{F}_{i+k}^\infty}|P(A\cap B)-P(A)P(B)|,$
and $\mathcal{F}_i^j$ is the $\sigma$-field generated by $\{\bw_t:i\leq t\leq j\}$ or $\{(\bx_{2t},\bff_{2t}):i\leq t\leq j\}$.  We use $c$ or $C$ as a finite positive generic constant below.

\subsection{Asymptotic Properties When $p$ is Fixed, but $n\rightarrow\infty$}
We begin with some assumptions.
\begin{assumption}
(i) The process $\{\bw_t\}$ is $\alpha$-mixing with the mixing coefficient satisfying the condition $\alpha_p(k)\leq \exp(-cn^{\gamma_1})$ for some constants $c>0$ and $\gamma_1\in(0,1]$. \\
(ii) For any $x>0$, $\sup_{t>0,1\leq i\leq r_1}P(|w_{it}|>x)\leq c\exp(-cx^{\gamma_2})$ for constants $c>0$,  $\gamma_2\in(0,2]$.\\
(iii) There exists a Gaussian process $\bW(\bt)=(W_1(t_1),...,W_{r_1}(t_{r_1}))'$ such that as $n\rightarrow \infty$,
$\bS_n^{r_1}(\bt)\overset{J_1}{\Longrightarrow}\bW(\bt)\,\, \text{on}\,\, D_{r_1}[0,1],$
where $\overset{J_1}{\Longrightarrow}$ denotes weak convergence under Skorokhod $J_1$ topology, and $\bW({\bf 1})$ has a positive definite covariance matrix $\bOmega=[\sigma_{ij}]$.
\end{assumption}
\begin{assumption}
The process $\{(\bx_{2t},\bff_{2t})\}$ is $\alpha$-mixing with the mixing coefficient satisfying the condition $\sum_{k=1}^\infty\alpha_p(k)^{1-2/\gamma}<\infty$ for some $\gamma>2$.
\end{assumption}
\begin{assumption}
(i) $E|f_{2,it}|^{2\gamma}<C$ and $E|\ve_{jt}|^{2\gamma}<C$ for $1\leq i\leq r_2$ and $1\leq j\leq v$, where  $\gamma$ is given in Assumption 2.\\
(ii) For any $\bh_1\in R^{r_1}$ and $\bh_2\in R^{v}$  with $\|\bh_1\|_2=\|\bh_2\|_2=1$, there exists a constant $C>0$ such that
$P(|\bh_1'\bff_{2t}|>x)\leq 2\exp(-Cx)$ and $P(|\bh_2'\bve_t|>x)\leq 2\exp(-Cx)$ for any $x>0$.
\end{assumption}
The mixing conditions in Assumptions 1(i) and 2  are standard for dependent random processes. See \cite{gaoetal2017} for a theoretical justification for VAR models. Assumption 1(ii) implies that the tail probabilities of linear combinations of $\bw_t$ decay exponentially fast, which is used to establish the Bernstein-type inequality of (\ref{bern}) together with Assumption 1(i) by Theorem 1 of \cite{merlevede2011}. The restrictions of $\gamma_1$ and $\gamma_2$ are introduced only for the presentation convenience and large $\gamma_1$ and $\gamma_2$ would only make the conditions therein stronger.
Assumption 1(iii) is also used in \cite{zhang-etal2019}. The conditions in Assumption 3(i) implies that $E|y_{it}|^{2\gamma}<C$ under the setting that $p$ is fixed, and Assumption 3(ii) suggests that $\bff_{2t}$ and $\bve_t$ are sub-exponential, which is a larger class of distributions than sub-Gaussian. The following theorem establishes the consistency of the estimated loading matrices $\wh\bA_1$, $\wh\bA_2$,  $\wh\bA_2\wh\bU_1$, its orthonormal complement $\wh\bA_2\wh\bV_1$, the matrix $\wh\bA_2\wh\bV_2$, and the extracted common factors $\wh\bA_1\wh\bx_{1t}$ and $\wh\bA_2\wh\bU_1\wh\bz_{2t}$. 
\begin{theorem}\label{tm1}
Suppose Assumptions 1--3 hold and $r_1$ and $r_2$ are known and fixed. Then, for fixed $p$,
\begin{equation}\label{Ai}
D(\mathcal{M}(\wh\bA_i),\mathcal{M}(\bA_i))=O_p(n^{-1}),\,\, \text{for}\,\, i=1,2, 
\end{equation}
\begin{equation}\label{A2U1}
D(\mathcal{M}(\wh\bA_2\wh\bU_{1}),\mathcal{M}(\bA_2\bU_{1}))=O_p(n^{-1/2}), D(\mathcal{M}(\wh\bA_2\wh\bV_{1}),\mathcal{M}(\bA_2\bV_{1}))=O_p(n^{-1/2}),
\end{equation}
\begin{equation}\label{A2V2}
D(\mathcal{M}(\wh\bA_2\wh\bV_{2}),\mathcal{M}(\bA_2\bV_{2}))=O_p(n^{-1/2}).
\end{equation}
As a result,
\begin{equation}\label{ext:ft}
\|\wh\bA_1\wh\bA_1'\by_t-\bA_1\bx_{1t}\|_2=O_p(n^{-1/2})\quad\text{and}\quad \|\wh\bA_2\wh\bU_1\wh\bz_{2t}-\bA_2\bU_1\bz_{2t}\|_2=O_p(n^{-1/2}).
\end{equation}
\end{theorem}
Similar results are also obtained in \cite{zhang-etal2019} when each component of $\by_t$ is $I(1)$. 
The standard $\sqrt{n}$-consistency is still achieved because the super-consistency of $\bA_i$ does not affect the second step asymptotically.
If we further assume that the largest $r_2$ eigenvalues of $\bM_2$ are distinct, 
$\bU_1$ can be uniquely defined and estimated if we ignore the signs of the columns, as described in the illustrations of Assumption 3 of \cite{gaotsay2019b} and Theorem 1 therein. Nevertheless, we just estimate the corresponding linear space spanned by its columns, and this will not alter the uniquenesses of $\wh\bA_1\wh\bA_1'\by_t$ and $\wh\bA_2\wh\bU_1\wh\bz_{2t}$ since any orthonormal rotations of the columns of $\wh\bA_1$ or   $\wh\bA_2\wh\bU_1$ will be  canceled out by its inverse associated with the rest matrix product, as illustrated by \cite{lamyao2012} for equation (2.3) therein.

The next theorem states that the proposed method in Section 2.3 can estimate $r_1$ and $r_2$ consistently.
\begin{theorem}
Under Assumptions 1--3, $ P(\wh r_1=r_1)\rightarrow 1$ and $P(\wh r_2=r_2)\rightarrow 1$ as $n\rightarrow\infty$.
\end{theorem}

\subsection{Asymptotic Properties When  $n\rightarrow\infty$ and $p\rightarrow\infty$}
{We consider the case when $p\rightarrow\infty$ with $p=O(n^c)$ for some constant $c$ which will be specified in the theorems below.}

\begin{assumption}
There exists a constant $\delta\in [0,1)$ such that $\sigma_1(\bL_1)\asymp$ ... $\asymp\sigma_{r_1}(\bL_1)\asymp p^{(1-\delta)/2}$, $\sigma_1(\bU_{22,1})\asymp$ ... $\asymp\sigma_{r_2}(\bU_{22,1})\asymp p^{(1-\delta)/2}$, $\sigma_1(\bU_{22,2})\asymp$ ... $\asymp\sigma_{K}(\bU_{22,2})\asymp p^{(1-\delta)/2}$, and $\sigma_{K+1}(\bU_{22,2})\asymp$ ... $\asymp\sigma_{v}(\bU_{22,2})\asymp 1$, where $0\leq K< v$ is an integer.
\end{assumption} 


\begin{assumption}
(i) The process $\{p^{-(1-\delta)/2}\bw_t\}$ is $\alpha$-mixing with the mixing coefficient satisfying the condition $\alpha_p(k)\leq \exp(-cn^{\gamma_1})$ for some constants $c>0$ and $\gamma_1\in(0,1]$. \\
(ii) For any $x>0$, $\sup_{t>0,1\leq i\leq r_1}P(|p^{-(1-\delta)/2}w_{it}|>x)\leq c\exp(-cx^{\gamma_2})$ for $c>0$,  $\gamma_2\in(0,2]$.\\
(iii) There exists a Gaussian process $\bW(\bt)=(W_1(t_1),...,W_{r_1}(t_{r_1}))'$ such that as $n\rightarrow \infty$,
$p^{-\frac{1-\delta}{2}}\bS_n^{r_1}(\bt)\overset{J_1}{\Longrightarrow}\bW(\bt)\,\,\text{on}\,\, D_{r_1}[0,1],$
where $\overset{J_1}{\Longrightarrow}$ denotes weak convergence under Skorohod $J_1$ topology, and $\bW({\bf 1})$ has a positive definite covariance matrix $\bOmega=[\sigma_{ij}]$.
\end{assumption}
\begin{assumption}
(i) For $\gamma$ given in Assumption 2, any $\bh\in R^{v}$ and $0<c_h<\infty$ with $\|\bh\|_2=c_h$, $E|\bh'\bve_t|^{2\gamma}<\infty$; (ii)  $\sigma_{\min}(\bR'\bV_2^{*}{'}\bU_1)\geq C_3$ for some constant $C_3>0$ and some half orthogonal matrix $\bR\in R^{(p-r_1-K)\times r_2}$ satisfying $\bR'\bR=\bI_{r_2}$, where $\sigma_{\min}$ denotes the minimum non-zero singular value of a matrix. 
\end{assumption}
 Assumption 4 is similar to Assumption 5 of \cite{gaotsay2019b} to quantify the strength of the factors and the noises, and Assumption 6 is the same as the one therein. Assumption 5 is similar to Assumption 1 in Section 3.1 and we take the strength of the nonstationary factors into account according to Assumption 4 and the decomposition of $\bL$ in Section 2.1. 
\begin{theorem}\label{tm3} 
Suppose Assumptions 2--6 hold and $r_1$ is known and fixed. If $r_2=o(\min(p^\delta,p^{1-\delta}))$ and $p=o(n^{1/(1+\delta)})$, then
\begin{equation}\label{Ai:p}
D(\mathcal{M}(\wh\bA_i),\mathcal{M}(\bA_i))=O_p(p^{1/2}n^{-1}),\,\, \text{for}\,\, i=1,2.
\end{equation}
Furthermore,
\begin{equation}\label{A2U1:p}
D(\mathcal{M}(\wh\bA_2\wh\bU_{1}),\mathcal{M}(\bA_2\bU_{1}))=O_p(\frac{p^{(1+\delta)/2}}{n^{1/2}}), D(\mathcal{M}(\wh\bA_2\wh\bV_{1}),\mathcal{M}(\bA_2\bV_{1}))
=O_p(\frac{p^{(1+\delta)/2}}{n^{1/2}}),
\end{equation}
\begin{equation}\label{A2V2:p}
D(\mathcal{M}(\wh\bA_2\wh\bV_{2}^{*}),\mathcal{M}(\bA_2\bV_{2}^{*}))=O_p(p^{(1+\delta)/2}n^{-1/2}).
\end{equation}
Consequently,
\begin{equation}\label{ext:ft:1}
p^{-1/2}\|\wh\bA_1\wh\bA_1'\by_t-\bA_1\bx_{1t}\|_2=O_p(p^{(1-\delta)/2}n^{-1/2}),
\end{equation}
\begin{equation}\label{ext:ft:2}
p^{-1/2}\|\wh\bA_2\wh\bU_1\wh\bz_{2t}-\bA_2\bU_1\bz_{2t}\|_2=O_p(p^{1/2}r_2^{1/2}n^{-1/2}+p^{-1/2}).
\end{equation}
\end{theorem}
From Theorem~\ref{tm3}, the convergence rate of $\mathcal{M}(\wh\bA_1)$ and $\mathcal{M}(\wh\bA_2)$ does not depend on the strength $\delta$ and coincides with  that in Theorem 3.3 of \cite{zhang-etal2019} if  assuming $r_1$ is small. Under this assumption and even with a slowly growing $r_2$, we can handle the cases when the dimension $p=o(n^{1/(1+\delta)})$ which is higher than the maximal rate of $p=o(n^{1/2-\tau})$ in \cite{zhang-etal2019} for some $0<\tau<1/2$. In addition, if $p=o(n^{1/(1+\delta)})$, all the remaining estimators are all consistent as shown in Theorem~\ref{tm3}. The next theorem states the consistency of the estimated $\wh r_1$ and $\wh r_2$ with additional constraints on $p$.

\begin{theorem}\label{tm4}
Let Assumptions 2--6 hold, $r_1$ be fixed and $r_2=o(\min(p^\delta,p^{1-\delta}))$.
(i) if $p=o(n^{1/(1+\delta)})$, $ P(\wh r_1=r_1)\rightarrow 1$ as $n\rightarrow\infty$; and
(ii) if $p^{1+\delta/2}n^{-1/2}\log(np)=o(1)$, $P(\wh r_2=r_2)\rightarrow 1$ as $n\rightarrow\infty$.
\end{theorem}

\section{Numerical Properties}
In this section, we study the finite-sample performance of the proposed methodology under the scenarios when $p$ is both small and large. 
As the dimensions of $\wh\bA_1$  and $\bA_1$ are not necessarily the same,
and  $\bL_1$ is not an orthogonal matrix in general, we first extend the discrepancy measure
in Equation (\ref{eq:D}) to a more general form below. Let $\bH_i$ be a
$p\times d_i$ matrix with rank$(\bH_i) = d_i$, and $\bP_i =
\bH_i(\bH_i'\bH_i)^{-1} \bH_i'$, $i=1,2$. Define
\begin{equation}\label{dmeasure}
\bar{D}(\bH_1,\bH_2)=\sqrt{1-
\frac{1}{\max{(d_1,d_2)}}\textrm{tr}(\bP_1\bP_2)}.
\end{equation}
Then $\bar{D} \in [0,1]$. Furthermore,
$\bar{D}(\bH_1,\bH_2)=0$ if and only if
either $\mathcal{M}(\bH_1)\subset \mathcal{M}(\bH_2)$ or
$\mathcal{M}(\bH_2)\subset \mathcal{M}(\bH_1)$, and  it is 1 if and only if
$\mathcal{M}(\bH_1) \perp \mathcal{M}(\bH_2)$.
If $d_1 = d_2=r$ and $\bH_i'\bH_i= \bI_r$, then 
$\bar{D}(\bH_1,\bH_2)$ reduces to that in Equation (\ref{eq:D}). 
We only present the simulation results 
for $k_0=2$ and $j_0=2$ in Equations (\ref{wy}) and (\ref{M2}), respectively, 
to save space because other choices of $k_0$ and $j_0$ produce similar results.\\

\subsection{Simulation}
{\bf Example 1.} Consider the models in (\ref{w:eq}) and (\ref{st:eq}) with common factors following
\[\bx_{1,t}=\bx_{1,t-1}+\bfeta_{1,t}\quad\text{and}\quad \bff_{2,t}=\bPhi \bff_{2,t-1}+\bfeta_{2,t},\]
where $\bfeta_{1,t}$ and $\bfeta_{2,t}$ are independent white noise processes. We set the true numbers of factors $r_1=2$ and $r_2=2$, the dimension $p=6, 10, 15, 20$, and the sample size $n=200, 500, 1000, 1500, 2000$. For each dimension $p$, we set the seed value to \texttt{1234} in \texttt{R} and generate first a matrix $\bM\in R^{p\times p}$ with each element independently drawn from $U(-2,2)$, then use the method and package of \cite{hoff2009} to simulate a random orthonormal matrix $\bA$ of Model (\ref{w:eq}) that follows a matrix-variate von Mises-Fisher distribution for a given matrix $\bM$.  The elements of $\bU_{22,1}$ and $\bU_{22,2}$ are drawn independently from $U(-1,1)$, and the elements of $\bU_{22,2}$ are then divided by $\sqrt{p}$ to balance the accumulated variances of $f_{2,it}$ and $\ve_{it}$ for each component of $\bx_{2,t}$. $\bPhi$ is a diagonal matrix with its diagonal elements being drawn independently from $U(0.5,0.9)$. For each realization, $\bve_t\sim N(0,\bI_v)$, $\bfeta_{1,t}\sim N(0,\bI_{r_1})$ and $\bfeta_{2,t}\sim N(0,\bI_{r_2})$,  we use $500$ replications for each $(p,n)$ configuration. For the estimation of $\wh r_1$, we use `a' to denote using the average of sample autocorrelations and 
`a$^*$'  the average of the absolute autocorrelations as described in Section 2.3. We denote `w$^*$' and `w'  the method of white noise test with and without the reordering procedure 
discussed in Section 2.3, respectively. `a$^*$w$^*$' is the estimation of $r_2$ using `w$^*$' based on the estimated $r_1$ by `a$^*$', and `aw' is similarly defined. 

\begin{figure}
\begin{center}
\subfigure[]{\includegraphics[width=0.325\textwidth]{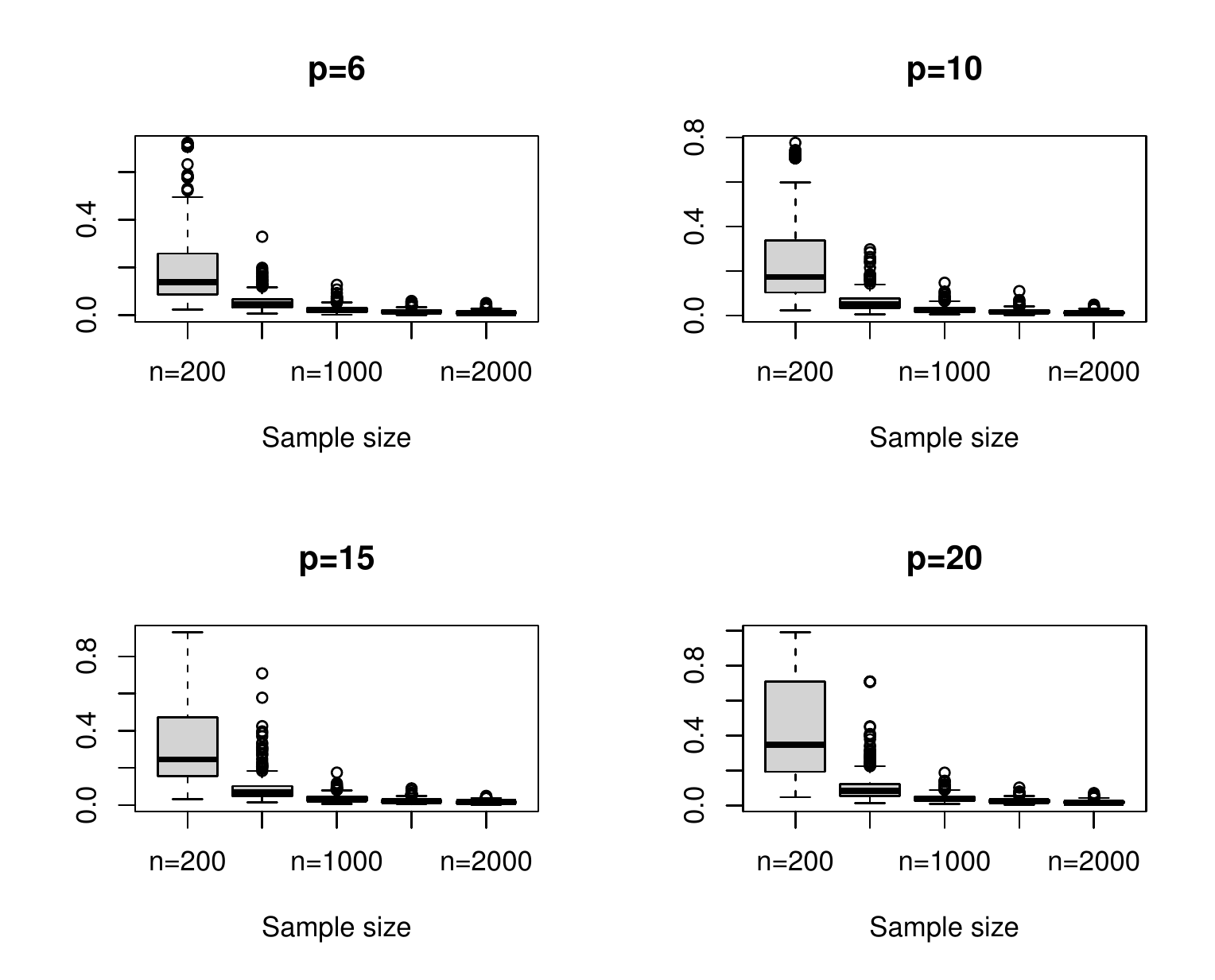}}
\subfigure[]{\includegraphics[width=0.325\textwidth]{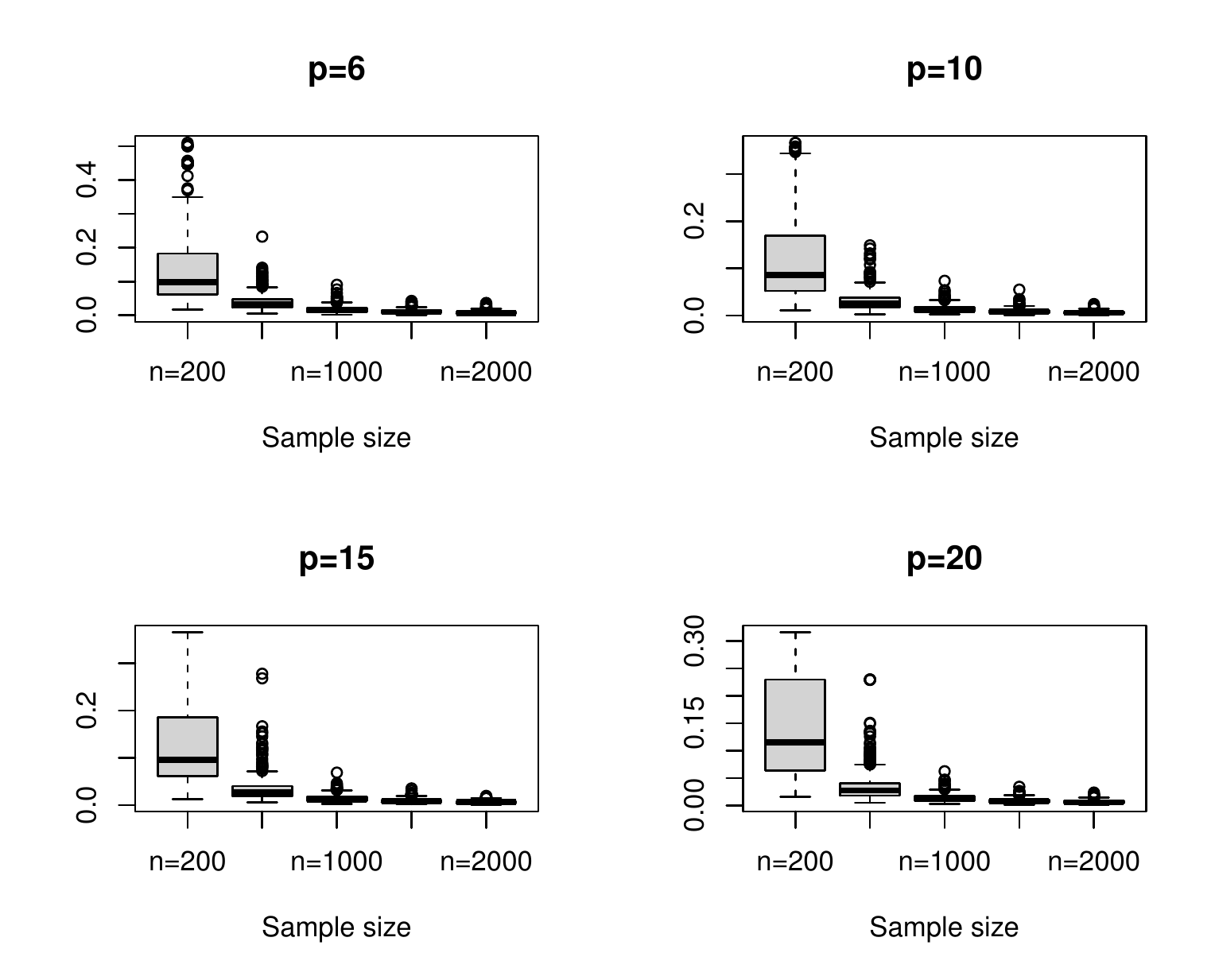}}
\subfigure[]{\includegraphics[width=0.325\textwidth]{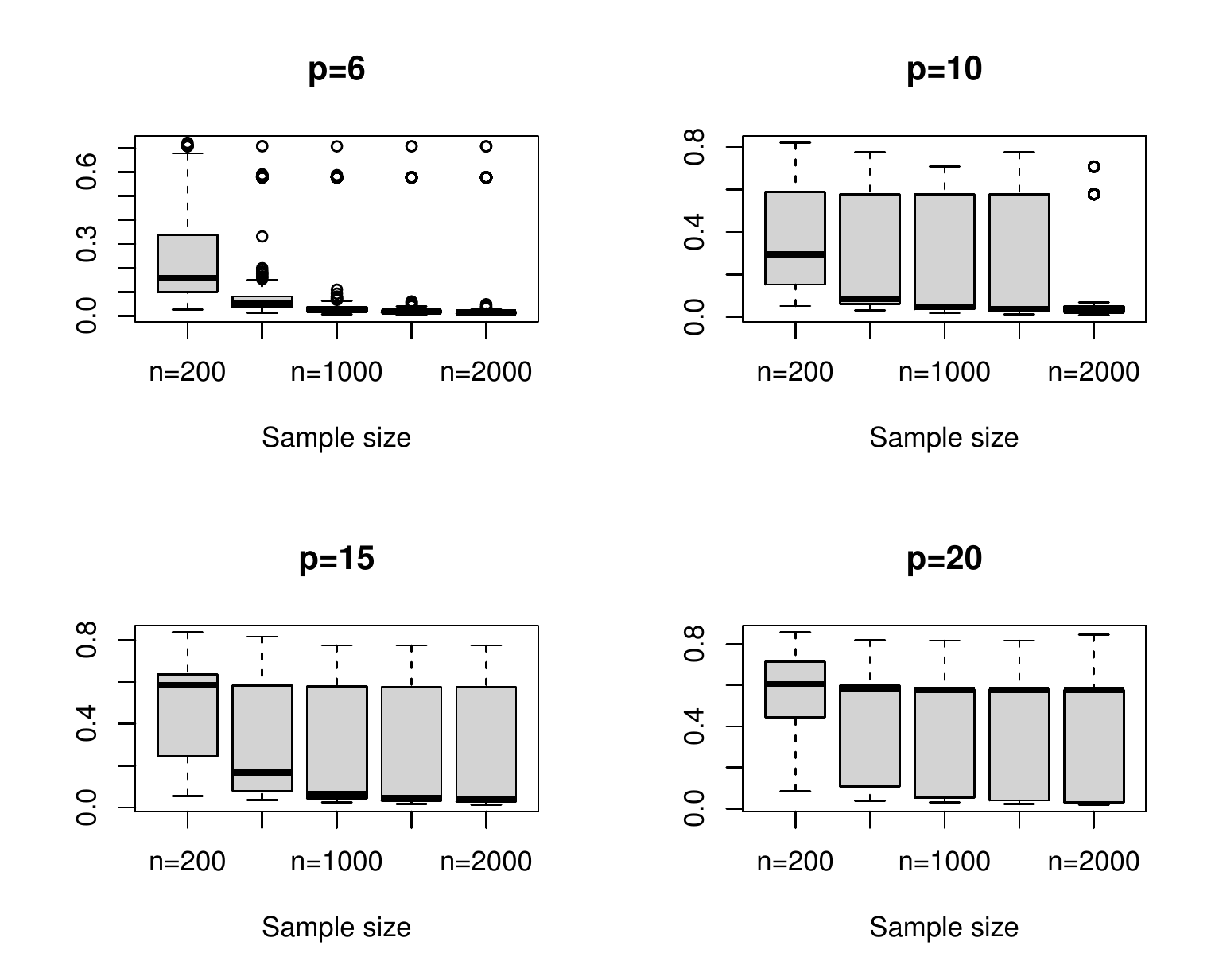}}
\caption{Estimation results of Example 1 when $p$ is relatively small and 
$r_1=r_2=2$: (a) Boxplots of $\bar{D}(\wh\bA_1,\bA_1)$; (b) Boxplots of $\bar{D}(\wh\bA_2,\bA_2)$; (c) Boxplots of $\bar{D}(\wh\bA_2\wh\bU_1,\bA_2\bU_{22,1})$. The sample sizes used are $200, 500, 1000, 1500, 3000$, and the results are based on $500$ iterations.}\label{fig1}
\end{center}
\end{figure}

We first study the performance of estimating the numbers of unit-root and stationary factors. We use $l=3$ and $m=10$ in (\ref{sm}) and $c_0=0.3$ to determine $r_1$. Since $p$ is relatively small compared to the sample size $n$, for each iteration, we use Ljung-Box test statistics $Q(m)$ with 
$m$ = 10 to determine the number of  common factors $r_2$. 
The empirical probabilities $P(\wh r_1=r_1)$, $P(\wh r_2=r_2)$, and $P(\wh r_1+\wh r_2=r_1+r_2)$ are reported in Table~\ref{Table1} for different methods. From the table, we see that, for a given $p$, the performance of all methods improves as the sample size increases. 
To detect the number of unit roots, the proposed method that uses average of absolute autocorrelations performs better than the one without using the absolute values when the sample size $n$ is not large. Also, the proposed method with reordering procedure also outperforms the one without reordering, especially when $n$ is relatively small. On the other hand, for a given $n$ and method, the empirical probability of correct specification 
decreases as $p$ increases, especially for the determination of $r_2$. This is understandable 
because it is harder to determine the correct number of stationary factors when the dimension increases and the errors in the  white noise testing procedure accumulates. In addition, when the sample size is small and the dimension is relatively high (e.g., $n=200$ and 
$p > 10$), the empirical probabilities $P(\wh r_1=r_1)$ and $P(\wh r_2=r_2)$ are not satisfactory, but the total number of factors ($r_1+ r_2$) can still be estimated reasonably well if the dimension is low, say $p \leq 10$. Finally,  for large $n$ (say, $n \geq 500$) and the low dimension $p$ considered in the simulation, the impact 
of modifications discussed in Section 2.3 is small. This is not surprising as the serial dependence 
of the stationary models used is positive and decays quickly. 
Overall, the estimation of $r_1$ performs well when the sample size is sufficiently large, 
and the Ljung-Box test to determine $r_2$ works well for the 
case of small dimension (e.g., $p\leq 10$). However, when $p$ is larger (e.g., $p=15, 20$), the Ljung-Box test statistic tends to overestimate the number of stationary factors, implying that we can still keep sufficient information of the original process $\by_t$. 
To illustrate the consistency in estimating the loading matrices of the proposed methods that use  absolute autocorrelations and the reordering procedure in white noise testing, we present the boxplots of $\bar{D}(\wh\bA_1,\bA_1)$, $\bar{D}(\wh\bA_2,\bA_2)$ and  $\bar{D}(\wh\bA_2\wh\bU_1,\bA_2\bU_{22,1})$ in Figure~\ref{fig1}(a), (b) and (c), respectively, where $\bar{D}(\cdot,\cdot)$ is defined in (\ref{dmeasure}). From Figure~\ref{fig1}, for each dimension $p$, the discrepancy decreases as the sample size increases and this is in agreement with the theory. 
The plots also show that, as expected, 
the mean discrepancy increases as the dimension $p$ increases. 

\begin{table}
 \caption{Empirical probabilities $P(\hat{r}_1=r)$, $P(\hat{r}_2=r_2)$ and $P(\wh r_1+\wh r_2 =r)$ of various $(p,n)$ configurations 
 for the model of Example 1 with $r_1=2$, $r_2=2$ and $r=r_1+r_2=4$, where $p$ and $n$ are the dimension and the sample size, respectively. `a'  denotes the method of using the average of sample autocorrelations and 
`a$^*$'  the average of absolute sample autocorrelations,  `w$^*$' and `w'  the method of white noise test with and without the reordering procedure of Section 2.3, respectively. `a$^*$w$^*$' is the estimation of $r_2$ using `w$^*$' based on the estimated $r_1$ by `a$^*$', and `aw' is similarly defined. $500$ iterations are used.} 
          \label{Table1}
\begin{center}
 \setlength{\abovecaptionskip}{0pt}
\setlength{\belowcaptionskip}{3pt}
\scriptsize
\begin{tabular}{c|c|r|rrrrr}
\hline
&&&\multicolumn{5}{c}{$n$}\\
$p$&EP&Methods&$200$&$500$&$1000$&$1500$&$3000$\\
\hline
&$P(\wh r_1=r_1)$&a$^*$(a)&0.874(0.682)&1(0.996)&1(1)&1(1)&1(1)\\
$6$&$P(\wh r_2=r_2)$& a$^*$w$^*$(aw)&0.788(0.604)&0.902(0.898)&0.906(0.906)&0.908(0.908)&0.914(0.914)\\
&$P(\wh r_1+\wh r_2=r)$&a$^*$w$^*$(aw)&0.908(0.904)&0.902(0.902)&0.906(0.906)&0.908(0.906)&0.914(0.914)\\
\hline
&$P(\wh r_1=r_1)$&a$^*$(a)&0.844(0.690)&1(1)&1(1)&1(1)&1(1)\\
$10$&$P(\wh r_2=r_2)$&a$^*$w$^*$(aw)&0.606(0.512)&0.740(0.740)&0.732(0.732)&0.726(0.726)&0.762(0.762)\\
&$P(\wh r_1+\wh r_2=r)$&a$^*$w$^*$(aw)&0.716(0.724)&0.740(0.740)&0.732(0.732)&0.726(0.726)&0.762(0.762)\\
   \hline 
   &$P(\wh r_1=r_1)$&a$^*$(a)&0.780(0.656)&0.996(0.994)&1(1)&1(1)&1(1)\\
$15$&$P(\wh r_2=r_2)$&a$^*$w$^*$(aw)&0.420(0.338)&0.544(0.540)&0.586(0.586)&0.592(0.592)&0.562(0.562)\\
&$P(\wh r_1+\wh r_2=r)$&a$^*$w$^*$(aw)&0.524(0.520)&0.544(0.544)&0.586(0.586)&0.592(0.592)&0.562(0.562)\\
  \hline 
   &$P(\wh r_1=r_1)$&a$^*$(a)&0.678(0.526)&0.988(0.982)&1(1)&1(1)&1(1)\\
$20$&$P(\wh r_2=r_2)$&a$^*$w$^*$(aw)&0.286(0.226)&0.390(0.390)&0.420(0.420)&0.434(0.434)&0.482(0.482)\\
&$P(\wh r_1+\wh r_2=r)$&a$^*$w$^*$(aw)&0.406(0.396)&0.398(0.400)&0.420(0.420)&0.434(0.434)&0.482(0.482)\\
\hline
\end{tabular}
  \end{center}
\end{table}

\begin{figure}
\begin{center}
\subfigure[]{\includegraphics[width=0.45\textwidth]{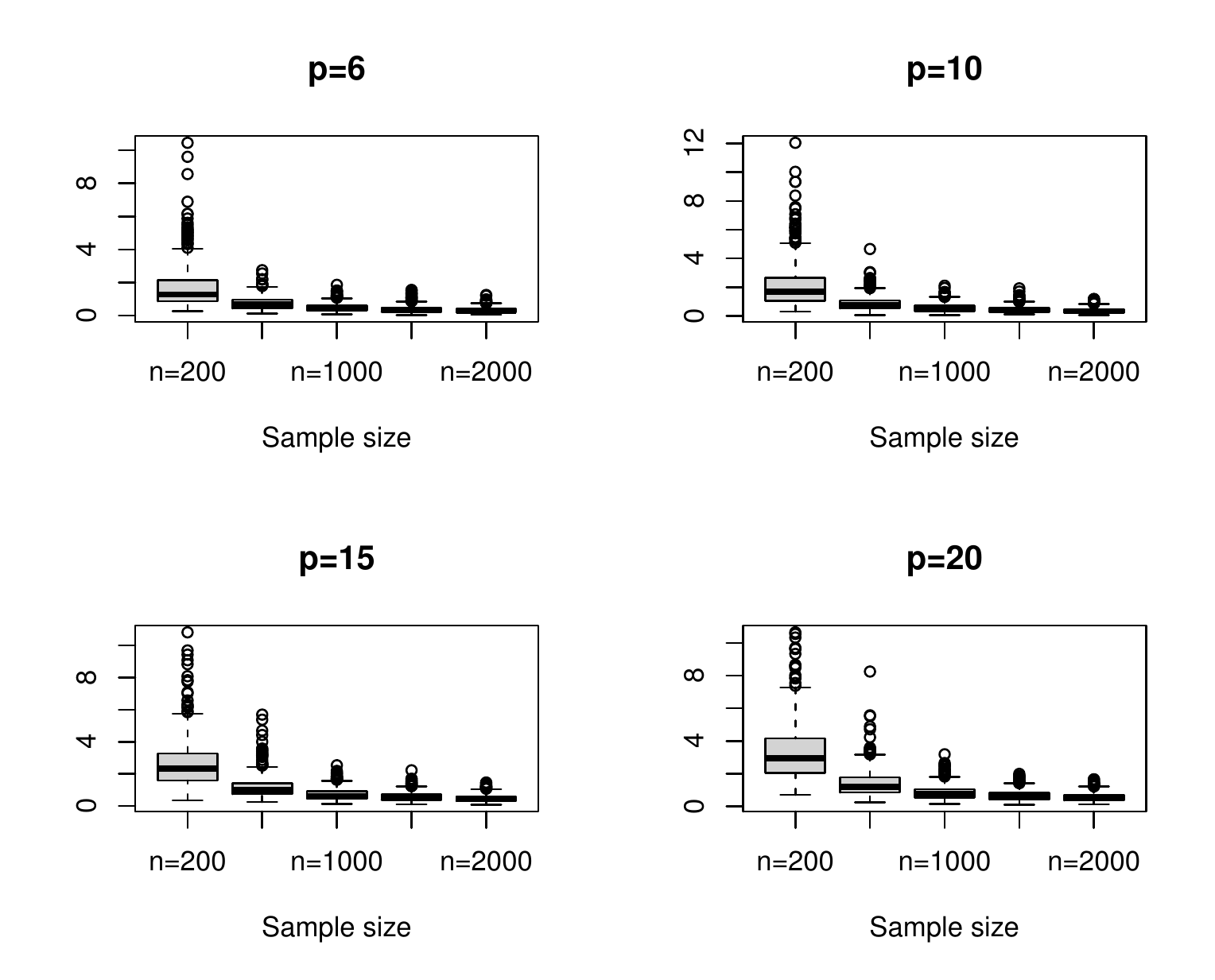}}
\subfigure[]{\includegraphics[width=0.45\textwidth]{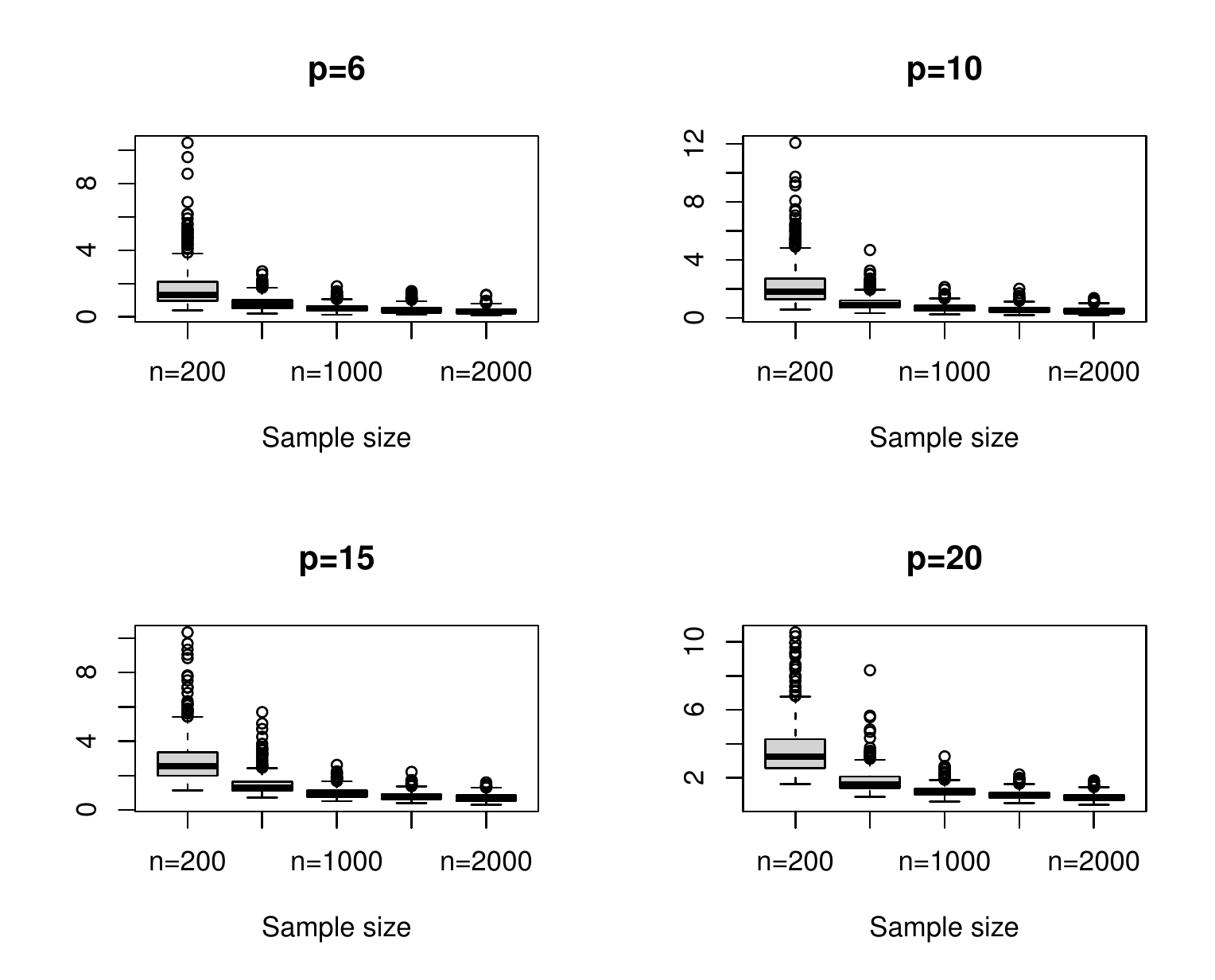}}
\caption{Estimation accuracies of Example 1 
when $p$ is relatively small and $r_1=r_2=2$: 
(a) Boxplots of the RMSE$_1$ defined in (\ref{rmse:psm}); (b) Boxplots of the RMSE$_2$ defined in (\ref{rmse:psm}). The sample sizes used are $200, 500, 1000, 1500, 3000$, and the results are based on $500$ iterations.}\label{fig2}
\end{center}
\end{figure}

Next, for each $(p,n)$ configuration, we study the root-mean-square errors (RMSEs):
\begin{equation}\label{rmse:psm}
\text{RMSE}_1=[\frac{1}{n}\sum_{t=1}^n\|\wh\bA_1\wh\bx_t-\bA_1\bx_{1t}\|_2^2]^{1/2},\text{RMSE}_2=[\frac{1}{n}\sum_{t=1}^n\|\wh\bA_2\wh\bU_1\wh\bz_{2t}-\bA_2\bU_{22,1}\bff_{2t}\|_2^2]^{1/2},
\end{equation} 
which quantify the accuracy in estimating the common factor processes.  
Boxplots of the RMSE$_1$ and RMSE$_2$ are shown in Figure~\ref{fig2}(a)-(b), respectively. From the plots, we  observe a clear pattern that, as the sample size increases, the RMSEs decrease for a given $p$, which is consistent with the results of Theorem \ref{tm1}. 
Overall, the one-by-one testing procedure works well when the dimension is small, and the RMSEs  decrease when the sample size increases, even though the performance of the 
Ljung-Box test may deteriorate due to the overestimation of the number of the 
stationary common factors for higher dimension  $p$.

\noindent{\bf Example 2.} In this example, we consider Models (\ref{w:eq}) and (\ref{st:eq}) with $\bx_{1t}$ and $\bff_{2t}$ 
being the same as those of Example 1. We set the true numbers of factors  $r_1=4$, $r_2=6$ and the number of prominent components of the noise covariance $K=2$  as defined in Assumption 4.  The dimensions 
used are $p=50, 100, 300, 500$, and the sample sizes are $n=300, 500, 1000, 1500, 2000$. We consider two scenarios for $\delta$ defined in Assumption 4: $\delta=0$ and $\delta=0.5$. For each setting, since it is time consuming to simulate a random orthonormal matrix when the dimension is high, instead we first simulate a matrix $\bM\in R^{p\times p}$ with each element drawn from $U(-2,2)$ using the same seed value as Example 1, then perform the SVD on $\bM$, and choose the columns of $\bA$ as the left singular vectors of $\bM$ multiplied by $p^{(1-\delta)/2}$.  The elements of $\bU_{22,1}$ and $\bU_{22,2}$ are drawn independently from $U(-1,1)$, then we divide $\bU_{22,1}$ by $p^{\delta/2}$,  the first $K$ columns of $\bU_{22,2}$ by $p^{\delta/2}$ and the remaining $v-K$ columns by $p$ to satisfy Assumption 4. 
$\bPhi$, $\bve_t$, $\bfeta_{1,t}$ and $\bfeta_{2,t}$ are drawn similarly as those of Example 1. We also use $500$ replications in each experiment and denote the methods by `a', `a$^*$', `w', and w$^*$  as in Example 1.

We first study the performance of $S_i(l,m)$ in (\ref{sm}) to estimate the number of  unit-root series  and that of high-dimensional white-noise tests of \cite{gaotsay2019b} to estimate the number of stationary common factors with and without the re-ordering modification of Section 2.3. 
The choices of $m$, $c_0$ and $l$ are the same as those in Example 1. For simplicity, we only present the results of the $T(m)$ statistics there to estimate $r_2$, and 
the results for the other white-noise test are similar. 
When $(p-\wh r_1)\geq n$, we only keep the upper $\epsilon n$ components of $\wh\bxi_t=\wh\bW'\wh\bA_2'\by_t$ with $\epsilon=0.75$ in white-noise testing. The results are reported in Table~\ref{Table2}. 
From the table, we see that for each setting of $\delta$ and fixed $p$, the performance of all methods improves as the sample size increases except for some cases in which 
$P(\wh r_2=r_2)$ of $n=500$ is smaller than that of $n=300$ for $p=500$. This is understandable since the actual dimension used in the white noise testing for $n=300$ is $0.75n=225$, while $(500-\wh r_1)$ for $n=500$ is much larger. For the estimation of $r_1$, the proposed method using absolute autocorrelations fares better when the sample size $n$ is small, and it is comparable with the one without using the absolute values when $n$ is large for each $p$. When estimating $r_2$, the proposed method consisting the reordering procedure (a$^*$w$^*$) outperforms 
the other in most cases when $n$ is relatively small ($n=300, 500$) 
 and the two methods are comparable with each other when the sample size is large.
As expected, when the factor strength is 
strong ($\delta = 0$), most procedures work better. The table also shows that 
the white noise test needs further improvement in selecting the number of stationary 
common factor $r_2$ when $p \geq n$, but the reordering procedure provides substantial 
improvements over the one without re-ordering.
Overall, the performance of the proposed modeling procedure is satisfactory 
when the sample size is larger than the dimension.
\begin{table}
 \caption{Empirical probabilities $P(\hat{r}_1=r)$, $P(\hat{r}_2=r_2)$ and $P(\wh r_1+\wh r_2 =r)$ of various $(p,n)$ configurations 
 for the model of Example 2 with $r_1=4$, $r_2=6$ and $r=r_1+r_2=10$, where $p$ and $n$ are the dimension and the sample size, respectively. `a'  denotes the method of using the average of  autocorrelations and 
`a$^*$'  the average of absolute autocorrelations,  `w$^*$' and `w'  denote the method of white noise test with and without the reordering procedure of Section 2.3. `a$^*$w$^*$' is the estimation of $r_2$ using `w$^*$' based on the estimated $r_1$ by `a$^*$', and `aw' is similarly defined. $500$ iterations are used.} 
          \label{Table2}
\begin{center}
 \setlength{\abovecaptionskip}{0pt}
\setlength{\belowcaptionskip}{2pt}
\tiny
\begin{tabular}{c|c|c|r|rrrrr}
\hline
&&&&\multicolumn{5}{c}{$n$}\\
$\delta$&$p$&EP&Methods&$300$&$500$&$1000$&$1500$&$2000$\\
\hline
$0$&&$P(\wh r_1=r_1)$&a$^*$(a)&0.798(0.412)&0.968(0.926)&0.992(0.992)&0.996(0.996)&0.998(0.998)\\
&$50$&$P(\wh r_2=r_2)$&a$^*$w$^*$(aw)&0.626(0.380)&0.892(0.846)&0.928(0.928)&0.898(0.898)&0.924(0.924)\\
&&$P(\wh r_1+\wh r_2=r)$&a$^*$w$^*$(aw)&0.770(0.778)&0.920(0.920)&0.936(0.936)&0.902(0.902)&0.926(0.926)\\
\cline{2-9}
&&$P(\wh r_1=r_1)$&a$^*$(a)&0.740(0.442)&0.930(0.928)&0.982(0.984)&0.998(0.998)&0.988(0.988)\\
&$100$&$P(\wh r_2=r_2)$&a$^*$w$^*$(aw)&0.634(0.408)&0.864(0.856)&0.898(0.900)&0.930(0.930)&0.904(0.904)\\
&&$P(\wh r_1+\wh r_2=r)$&a$^*$w$^*$(aw)&0.832(0.832)&0.924(0.924)&0.916(0.916)&0.932(0.932)&0.916(0.916)\\
\cline{2-9}
&&$P(\wh r_1=r_1)$&a$^*$(a)&0.748(0.450)&0.892(0.922)&0.962(0.980)&0.970(0.974)&0.982(0.982)\\
&$300$&$P(\wh r_2=r_2)$&a$^*$w$^*$(aw)&0.336(0.154)&0.804(0.830)&0.874(0.894)&0.902(0.908)&0.896(0.896)\\
&&$P(\wh r_1+\wh r_2=r)$&a$^*$w$^*$(aw)&0.370(0.234)&0.896(0.896)&0.910(0.910)&0.926(0.926)&0.910(0.910)\\
\cline{2-9}
&&$P(\wh r_1=r_1)$&a$^*$(a)&0.690(0.446)&0.916(0.926)&0.980(0.986)&0.962(0.962)&0.984(0.984)\\
&$500$&$P(\wh r_2=r_2)$&a$^*$w$^*$(aw)&0.526(0.408)&0.764(0.172)&0.906(0.912)&0.892(0.894)&0.924(0.922)\\
&&$P(\wh r_1+\wh r_2=r)$&a$^*$w$^*$(aw)&0.702(0.706)&0.824(0.188)&0.926(0.926)&0.928(0.928)&0.936(0.936)\\
\hline
$0.5$&&$P(\wh r_1=r_1)$&a$^*$(a)&0.786(0.440)&0.962(0.950)&0.996(0.996)&0.996(0.996)&0.996(0.996)\\
&$50$&$P(\wh r_2=r_2)$&a$^*$w$^*$(aw)&0.596(0.368)&0.880(0.868)&0.930(0.930)&0.928(0.928)&0.886(0.886)\\
&&$P(\wh r_1+\wh r_2=r)$&a$^*$w$^*$(aw)&0.728(0.830)&0.914(0.916)&0.934(0.934)&0.932(0.922)&0.890(0.890)\\
\cline{2-9}
&&$P(\wh r_1=r_1)$&a$^*$(a)&0.732(0.438)&0.936(0.930)&0.990(0.994)&0.994(0.994)&0.990(0.990)\\
&$100$&$P(\wh r_2=r_2)$&a$^*$w$^*$(aw)&0.436(0.300)&0.818(0.818)&0.912(0.916)&0.932(0.932)&0.902(0.902)\\
&&$P(\wh r_1+\wh r_2=r)$&a$^*$w$^*$(aw)&0.594(0.612)&0.866(0.882)&0.922(0.922)&0.938(0.938)&0.912(0.912)\\
\cline{2-9}
&&$P(\wh r_1=r_1)$&a$^*$(a)&0.764(0.450)&0.904(0.920)&0.968(0.980)&0.968(0.974)&0.982(0.982)\\
&$300$&$P(\wh r_2=r_2)$&a$^*$w$^*$(aw)&0.128(0.012)&0.474(0.146)&0.796(0.708)&0.888(0.892)&0.896(0.892)\\
&&$P(\wh r_1+\wh r_2=r)$&a$^*$w$^*$(aw)&0.144(0.018)&0.536(0.144)&0.820(0.720)&0.914(0.910)&0.914(0.910)\\
\cline{2-9}
&&$P(\wh r_1=r_1)$&a$^*$(a)&0.708(0.450)&0.914(0.916)&0.982(0.986)&0.958(0.958)&0.982(0.982)\\
&$500$&$P(\wh r_2=r_2)$&a$^*$w$^*$(aw)&0.264(0.008)&0.262(0.014)&0.628(0.318)&0.788(0.686)&0.880(0.852)\\
&&$P(\wh r_1+\wh r_2=r)$&a$^*$w$^*$(aw)&0.380(0.012)&0.274(0.010)&0.634(0.312)&0.822(0.710)&0.888(0.858)\\
\hline
\end{tabular}
  \end{center}
\end{table}

\begin{figure}
\begin{center}
{\includegraphics[width=0.6\textwidth]{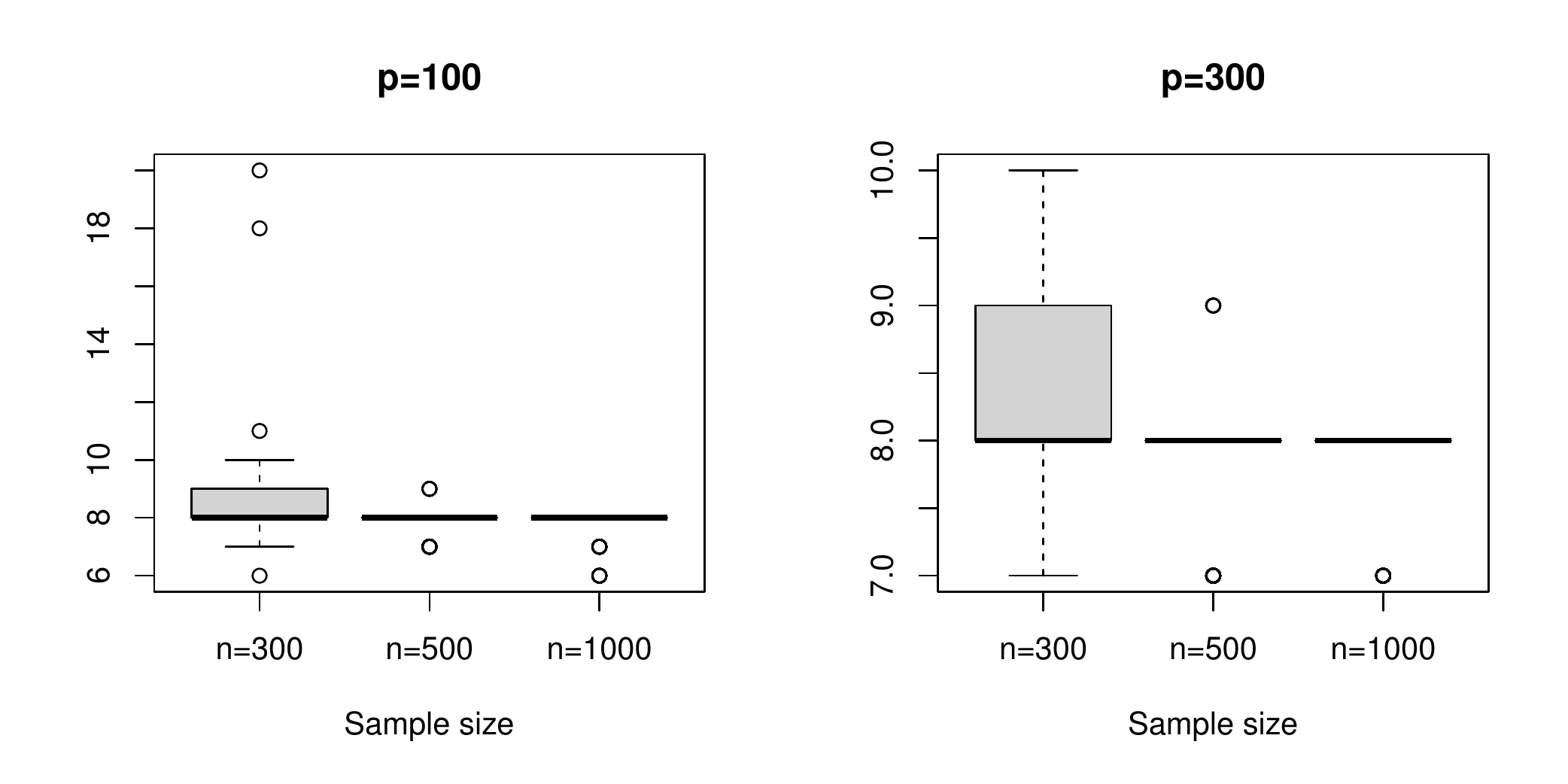}}
\caption{Boxplots of Lam and Yao (2012)'s ratio method in selecting the number of 
stationary common factors when  $r_1=4$, $r_2=6$, $K=2$. Left: $p=100$; 
Right: $p=300$ and the results are based on $500$ iterations.}\label{fig3a}
\end{center}
\end{figure}

To demonstrate the advantages of using  white-noise tests to select the number of stationary 
common factors with dynamic dependencies, we compare it with ratio-based method in \cite{lamyao2012}, which defines
$\wh r_2=\arg\min_{1\leq j\leq R}\left\{\wh\lambda_{j+1}/\wh\lambda_j\right\}$, where $\wh\lambda_1,...,\wh\lambda_{p-\wh r_1}$ denote the eigenvalues of $\wh\bM_2$ in (\ref{M2}) and we choose $R=(p-\wh r_1)/2$ suggested in their paper. Figure~\ref{fig3a} presents the boxplots of $\wh r_2$ for $p=100$ and $300$ with different sample sizes. 
The selection of $\wh r_2$ centers around 8, instead of the true value 6, 
indicating that the ratio-based method may fail to identify the correct number of stationary factors $r_2$ if the noise effect is diverging, while the high-dimensional white noise test continues to work well as shown in Table~\ref{Table2}.

\begin{figure}
\begin{center}
\subfigure[]{\includegraphics[width=0.325\textwidth]{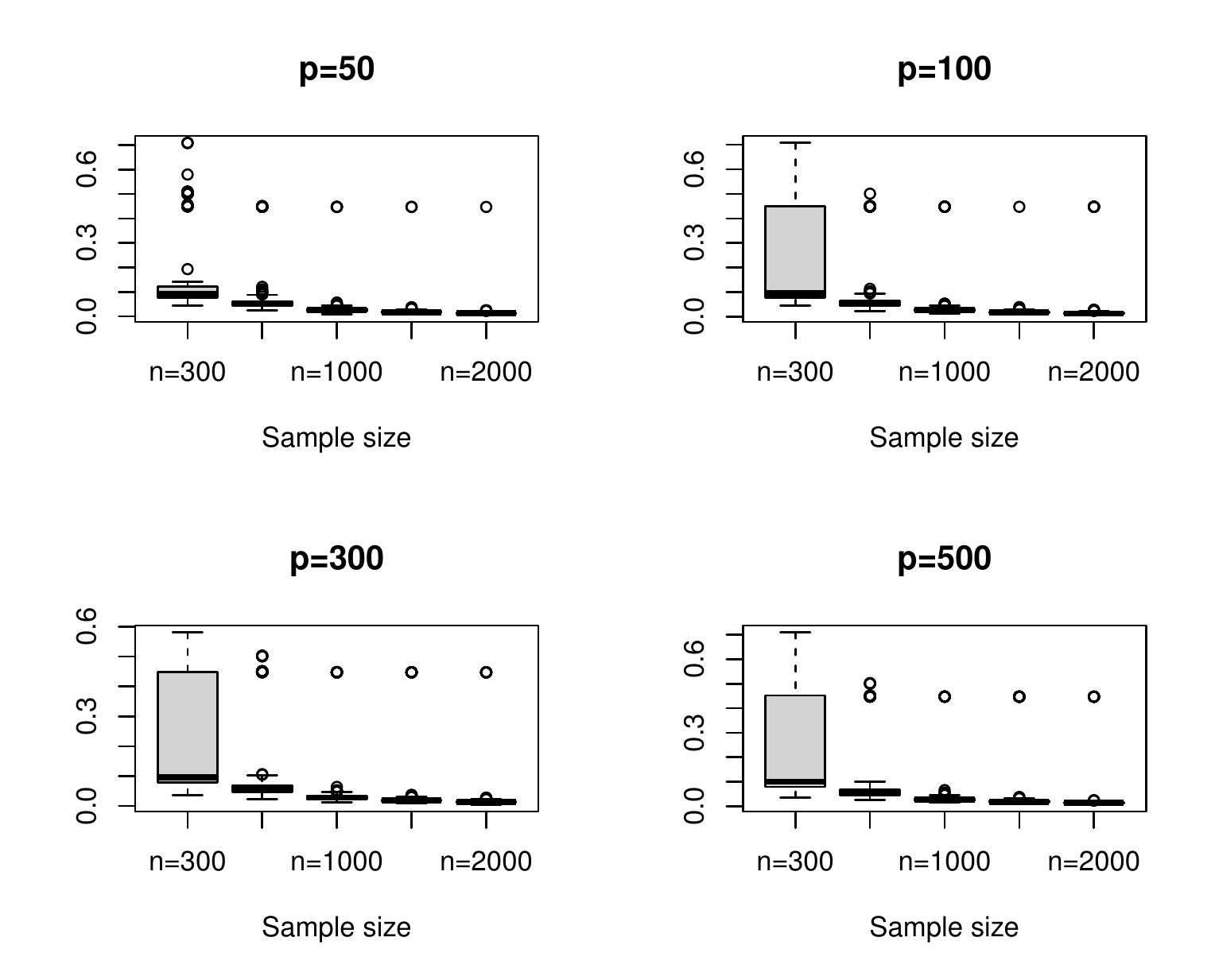}}
\subfigure[]{\includegraphics[width=0.325\textwidth]{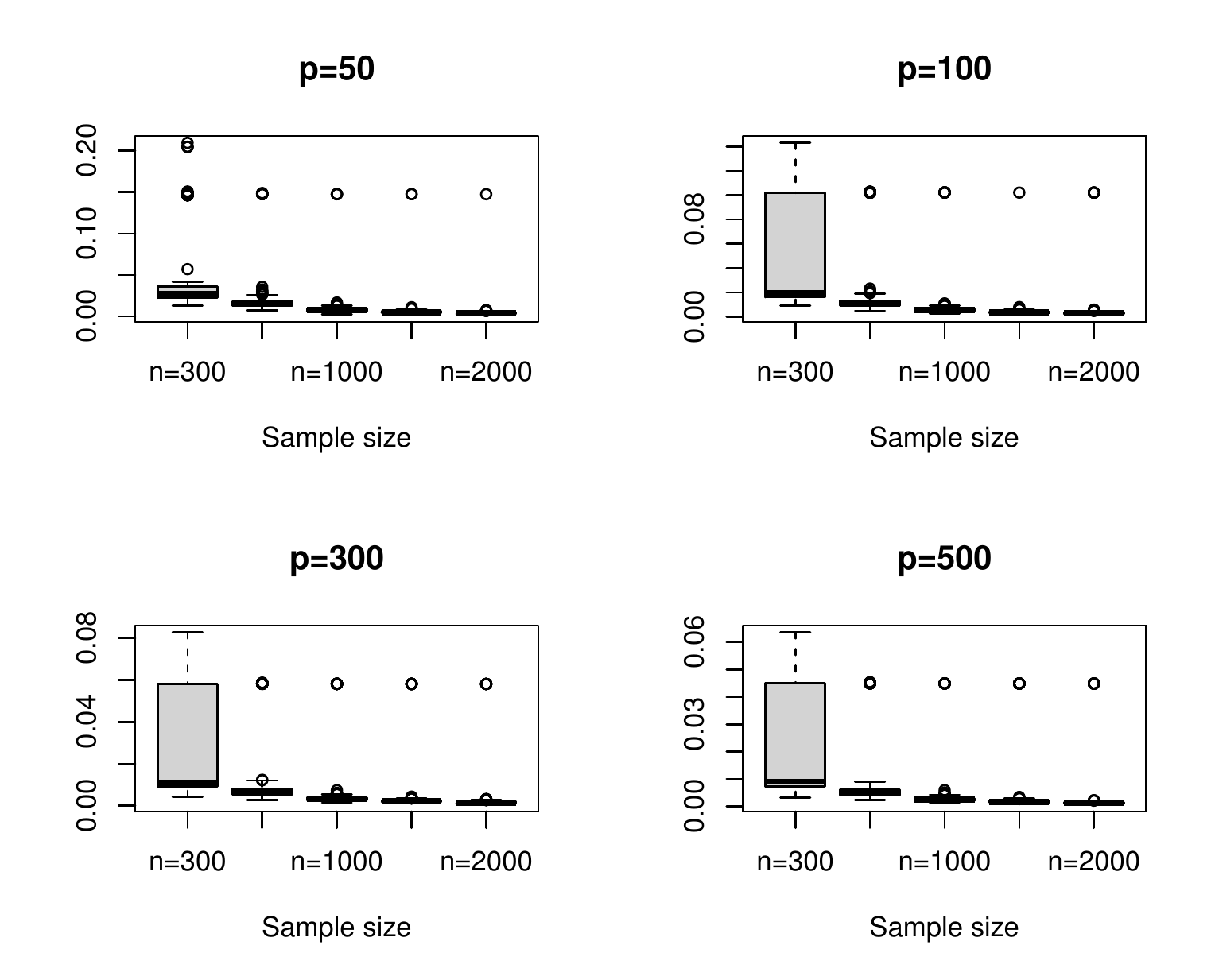}}
\subfigure[]{\includegraphics[width=0.325\textwidth]{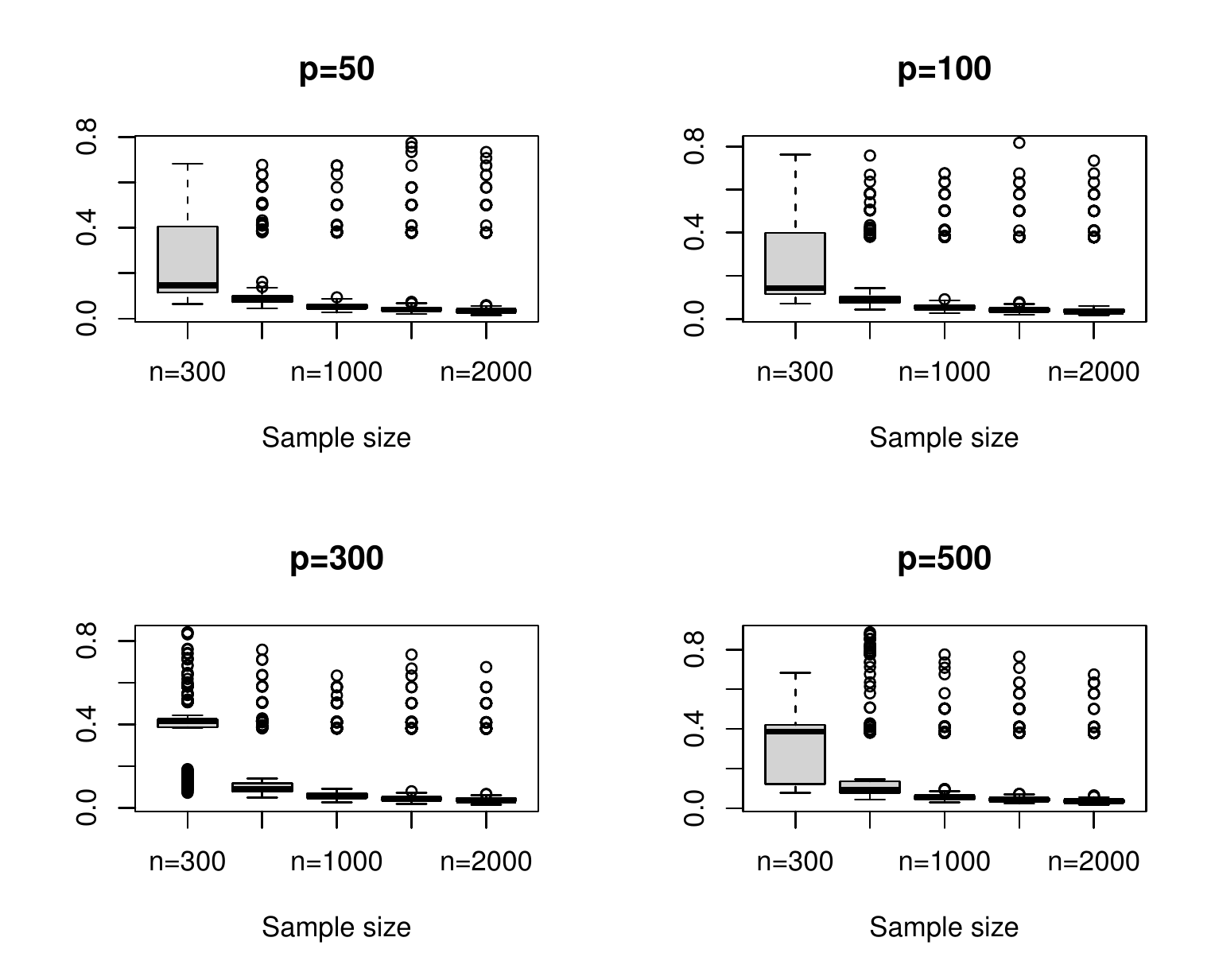}}
\caption{Estimation results of Example 2 when $p$ is relatively large, $r_1=4$, $r_2=6$, and 
$K=2$; (a) Boxplots of $\bar{D}(\wh\bA_1,\bA_1)$; (b) Boxplots of $\bar{D}(\wh\bA_2,\bA_2)$; 
(c) Boxplots of $\bar{D}(\wh\bA_2\wh\bU_1,\bA_2\bU_{22,1})$. The sample sizes used are $300, 500, 1000, 1500, 3000$, and the results are based on $500$ iterations.}\label{fig3}
\end{center}
\end{figure}

Next, we study the accuracy of the estimated loading matrices as that of Example 1 
using the methods a$^*$ and a$^*$w$^*$. 
The boxplots of $\bar{D}(\wh\bA_1,\bA_1)$, $\bar{D}(\wh\bA_2,\bA_2)$ and $\bar{D}(\wh\bA_2\wh\bU_1,\bA_2\bU_{22,1})$  are shown in Figure~\ref{fig3}(a)--(c), respectively. Similar patterns are also obtained  for the estimation of other matrices so we omit them here. 
From Figure~\ref{fig3}, 
there is a clear indication that the estimation 
accuracy of the loading matrix improves as the sample size increases even 
for moderately large $p$, 
which is in line with the asymptotic theory. The results also confirm that the proposed $S_i(l,m)$ and the white noise 
test that select $\wh r_1$ and $\wh r_2$, respectively, perform reasonably well even for large $p$. 

\begin{figure}
\begin{center}
\subfigure[]{\includegraphics[width=0.48\textwidth]{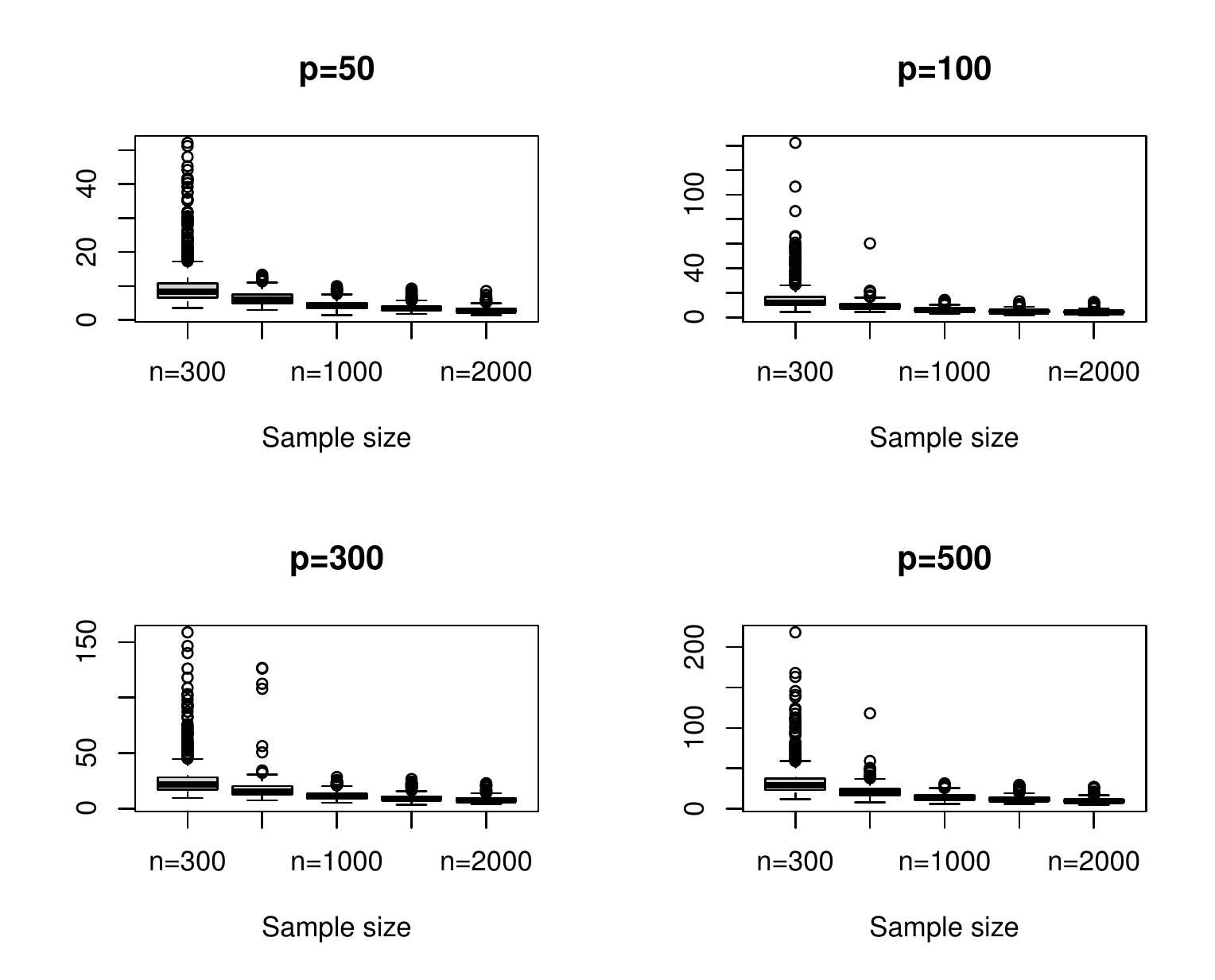}}
\subfigure[]{\includegraphics[width=0.48\textwidth]{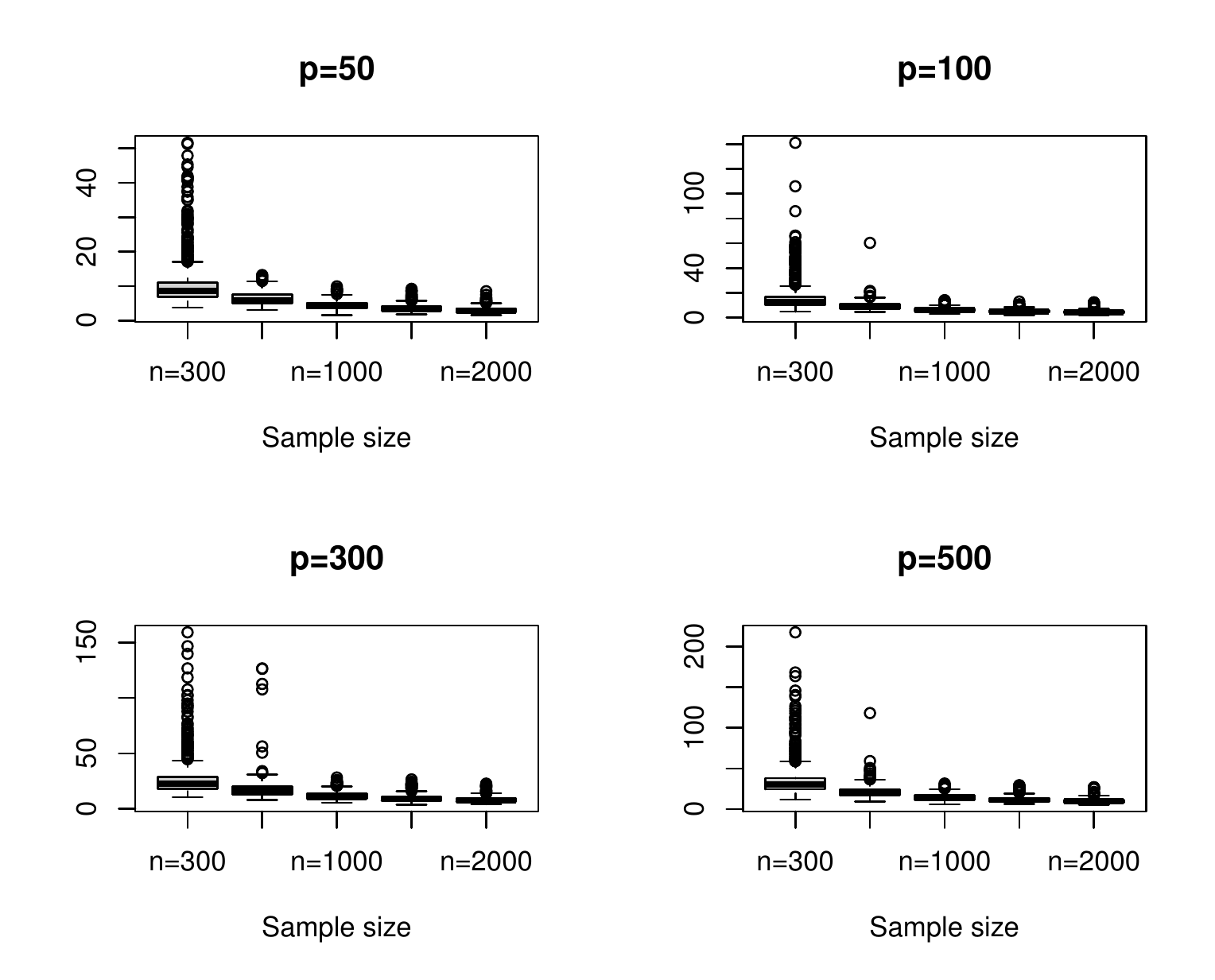}}
\caption{Estimation accuracies of Example 2 when $p$ is relatively large, 
$r_1=4$, $r_2=6$ and $K=2$: (a) Boxplots of the RMSE$_3$ defined in (\ref{rmse:plg}); 
(b) Boxplots of the RMSE$_4$ defined in (\ref{rmse:plg}). The sample sizes used are $300, 500, 1000, 1500, 2000$, and the results are based on $500$ iterations.}\label{fig4}
\end{center}
\end{figure}

For each $(p,n)$, we further define the RMSEs for high dimension as:{\small
\begin{equation}\label{rmse:plg}
\text{RMSE}_3=[\frac{1}{np}\sum_{t=1}^n\|\wh\bA_1\wh\bx_t-\bA_1\bx_{1t}\|_2^2]^{1/2},\text{RMSE}_4=[\frac{1}{np}\sum_{t=1}^n\|\wh\bA_2\wh\bU_1\wh\bz_{2t}-\bA_2\bU_{22,1}\bff_{2t}\|_2^2]^{1/2},
\end{equation} }
which quantify the accuracy in estimating the common factor processes.  
When the dimension is moderately large, the
boxplots of the RMSE$_3$ and RMSE$_4$ are shown in Figure~\ref{fig4}(a)-(b), respectively. From the plots, similar to Example 1, we  see a clear pattern that, as the sample size increases, the RMSEs decrease for a given $p$, which is consistent with the results of Theorem \ref{tm3}. 
Overall, the proposed method works well even for moderately high dimensions. 
This is especially so when the sample size is greater than the dimension.

\subsection{Real Data Analysis}
\begin{figure}
\begin{center}
{\includegraphics[width=0.8\textwidth]{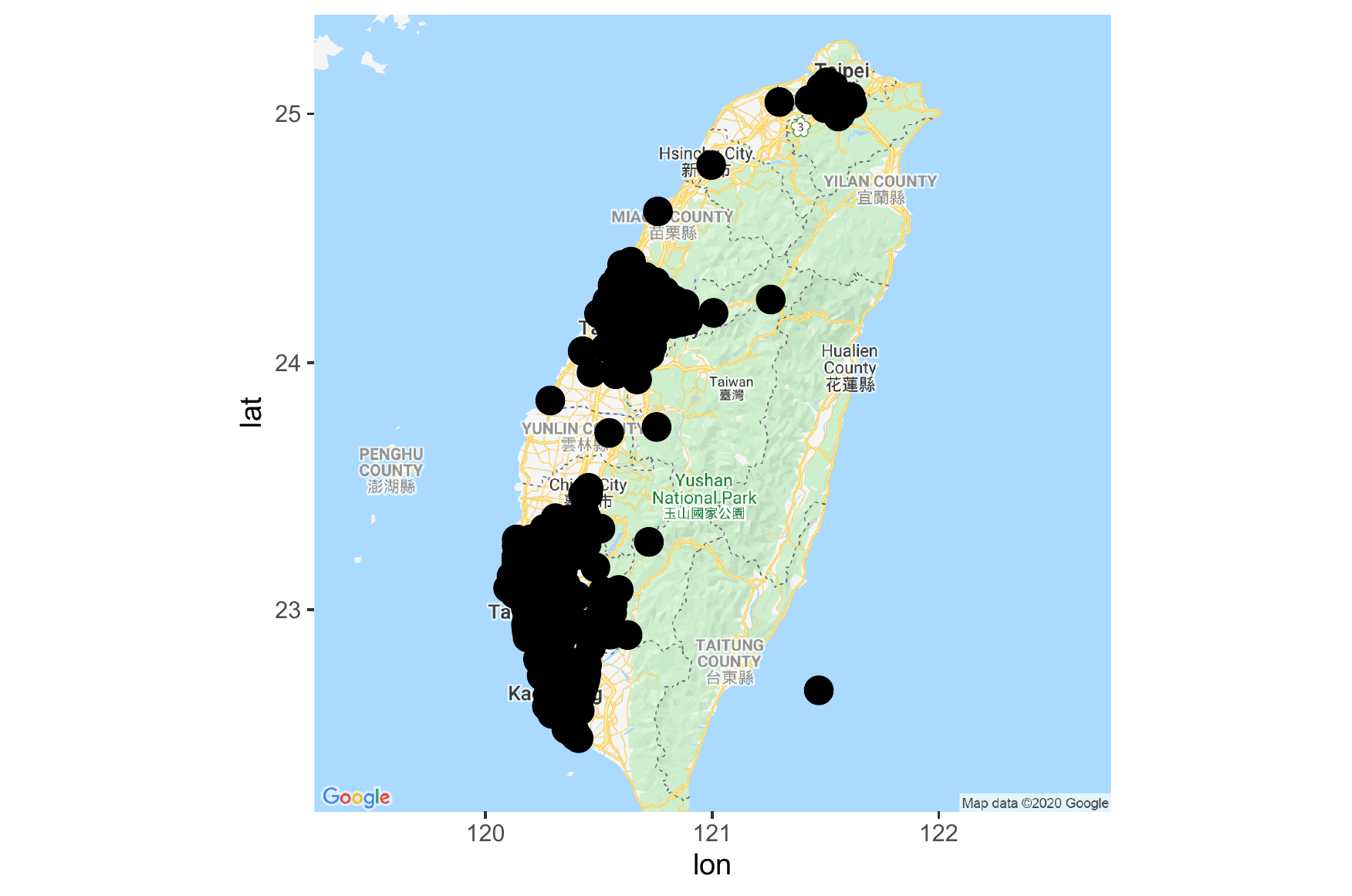}}
\caption{Locations (latitude vs. longitude) of the 508 AirBoxes across Taiwan of Example 3.}\label{TWNlocations}
\end{center}
\end{figure}

\begin{figure}
\begin{center}
{\includegraphics[width=0.8\textwidth]{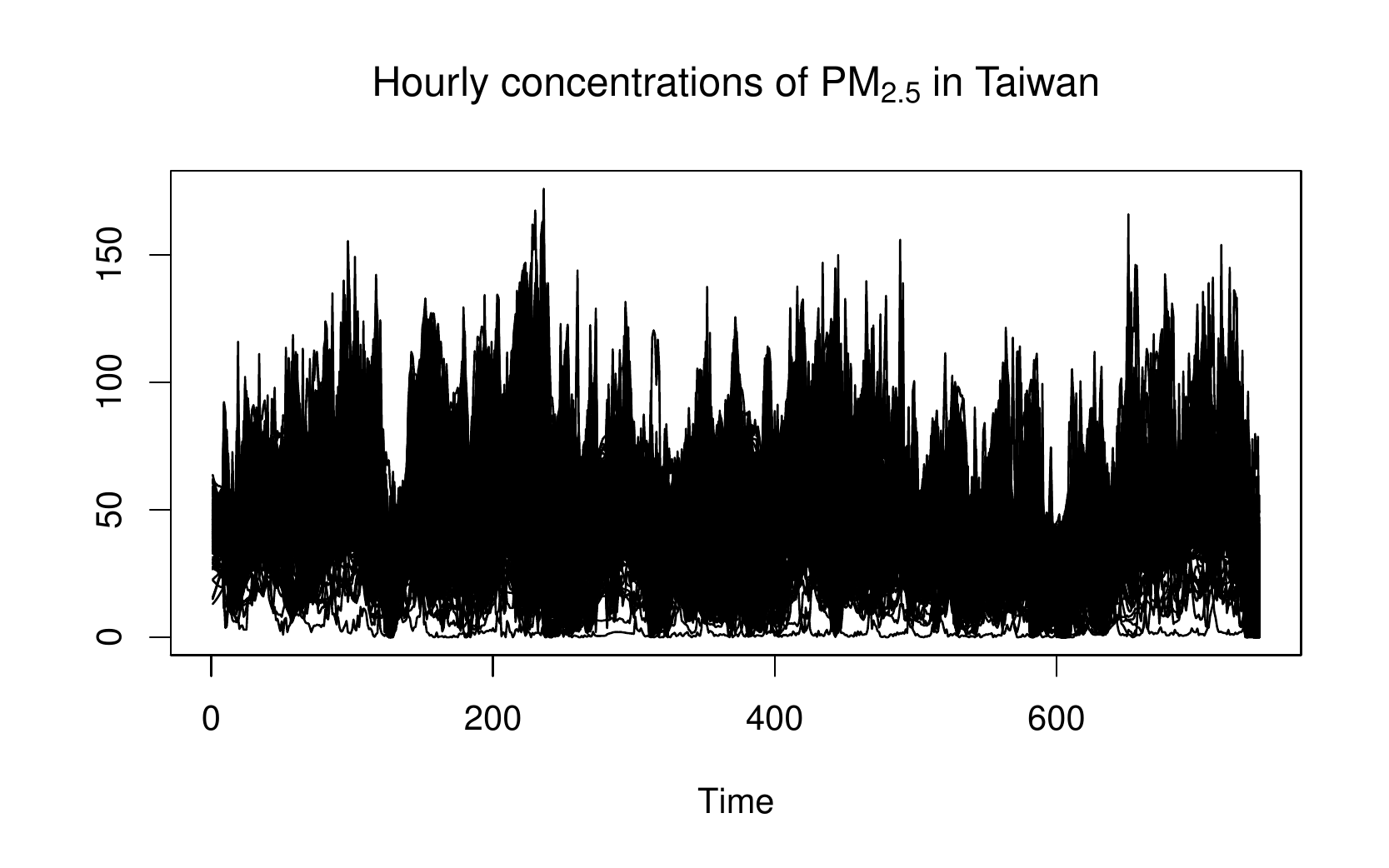}}
\caption{Time plots of hourly measurements of PM$_{2.5}$ of 508 locations  across Taiwan in March 2017 of Example 3.}\label{PM25plots}
\end{center}
\end{figure}

{\bf Example 3.} Consider the hourly measurements of PM$_{2.5}$ collected by 
Airboxes at 508 locations in Taiwan. The locations of the 508 locations are shown in Figure~\ref{TWNlocations}, which mainly consist of three clusters, signifying the major cities 
of Taiwan, namely Taipei, Taichung, Tainan, and Kaohsiung. The latter two cities are adjacent and 
form a large cluster. 
The isolated location outside of Taiwan  
denotes part of the Orchid Island of Taiwan. 
We apply our proposed method to the hourly measurements of PM$_{2.5}$ for March 2017 
with a total of 744 observations. The 508 time series are shown in Figure~\ref{PM25plots}.

We first applied the method of Section 2.3 with $k_0=2$ and found that $\wh r_1=3$, 
i.e., there are 3 unit-root 
processes. When applying the method, we choose $c_0=0.3$, $m=30$, and $l=3$. Similar results are also obtained by varying the values of $c_0$, $m$, and $l$ for several choices.  
For example, when $(c_0,m,l)=(0.3,15,2)$, $(0.2,15,3)$, and $(0.2,8,4)$, we obtain that $\wh r_1=3$, 3, and 1, respectively. 
It is true that these choices are subjective, but they are not unique to our empirical study. 
See, for instance, \cite{bai2004} in selecting the number of factors for different choices of $kmax$.
However, the key message of the detection is that there exist some unit-root common trends in 
the hourly PM$_{2.5}$ measurements. 

\begin{figure}
\begin{center}
{\includegraphics[width=0.62\textwidth]{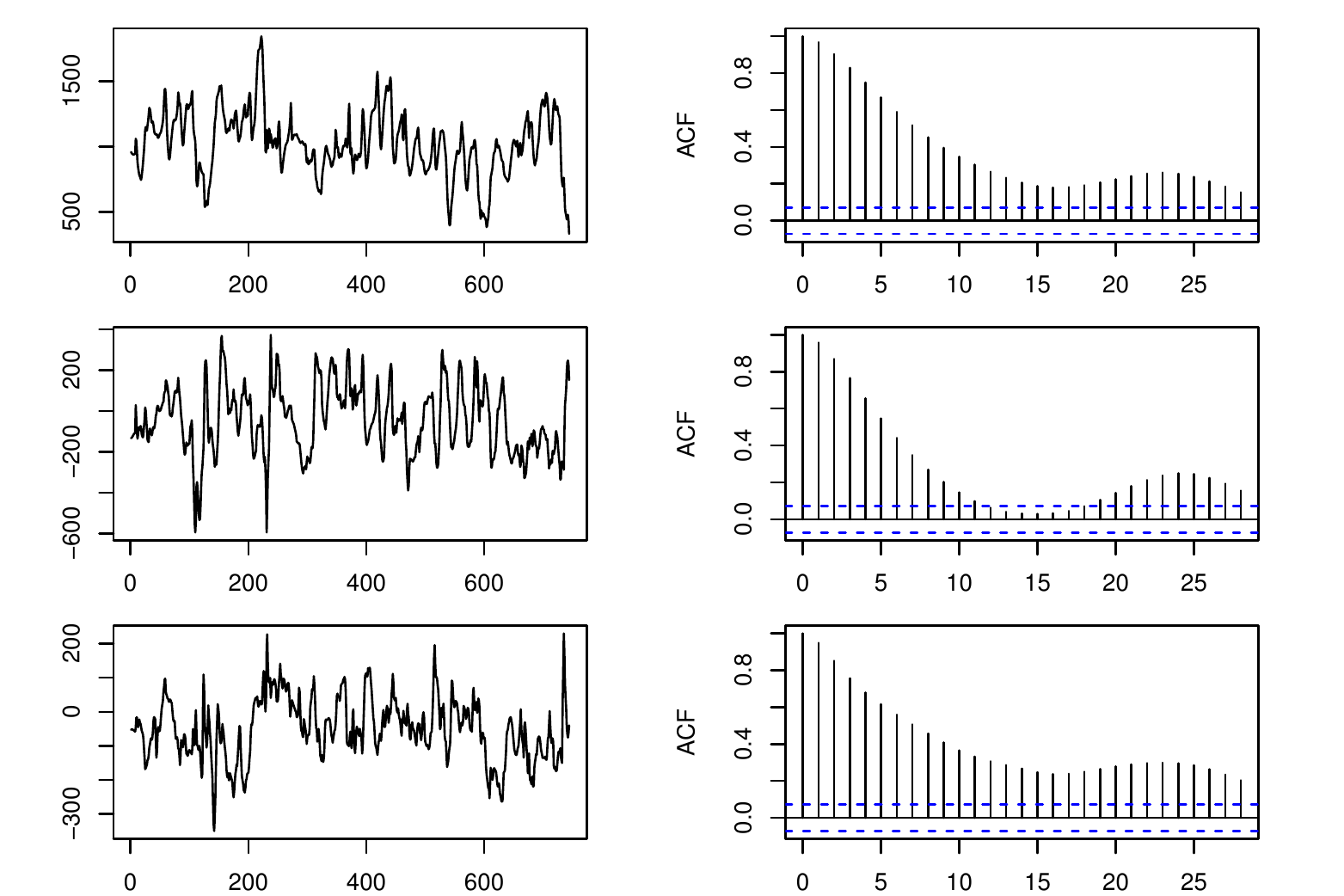}}
\caption{Time plots of the 3 estimated unit-root common trends by the proposed method and their sample ACFs of Example 3.}\label{PM25unitroot}
\end{center}
\end{figure}

The three recovered unit-root factors and their sample ACFs are shown in Figure~\ref{PM25unitroot}. 
From the plots, we see that these three unit-root factors capture most of the trends in the 
original data of Figure~\ref{PM25plots} and, as expected, their sample ACFs 
decay slowly. The extracted 505-dimensional stationary process is shown in
Figure~\ref{fig8}(a). 


\begin{figure}
\begin{center}
{\includegraphics[width=0.62\textwidth]{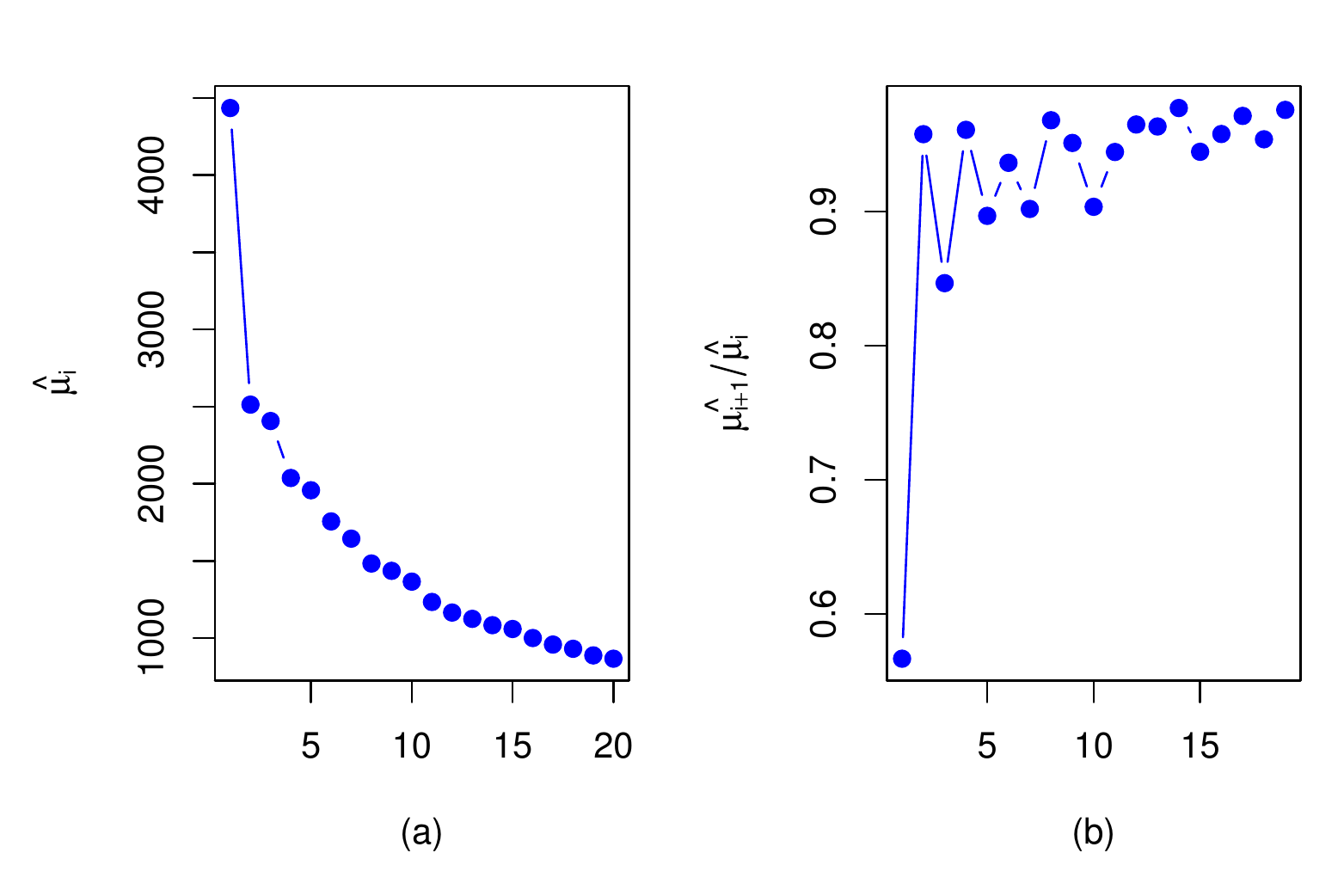}}
\caption{(a) The first 20 eigenvalues of $\wh\bS$; (b) The plot of ratios of consecutive eigenvalues of $\wh\bS$ .}\label{fig7}
\end{center}
\end{figure}

\begin{figure}
\begin{center}
{\includegraphics[width=0.6\textwidth]{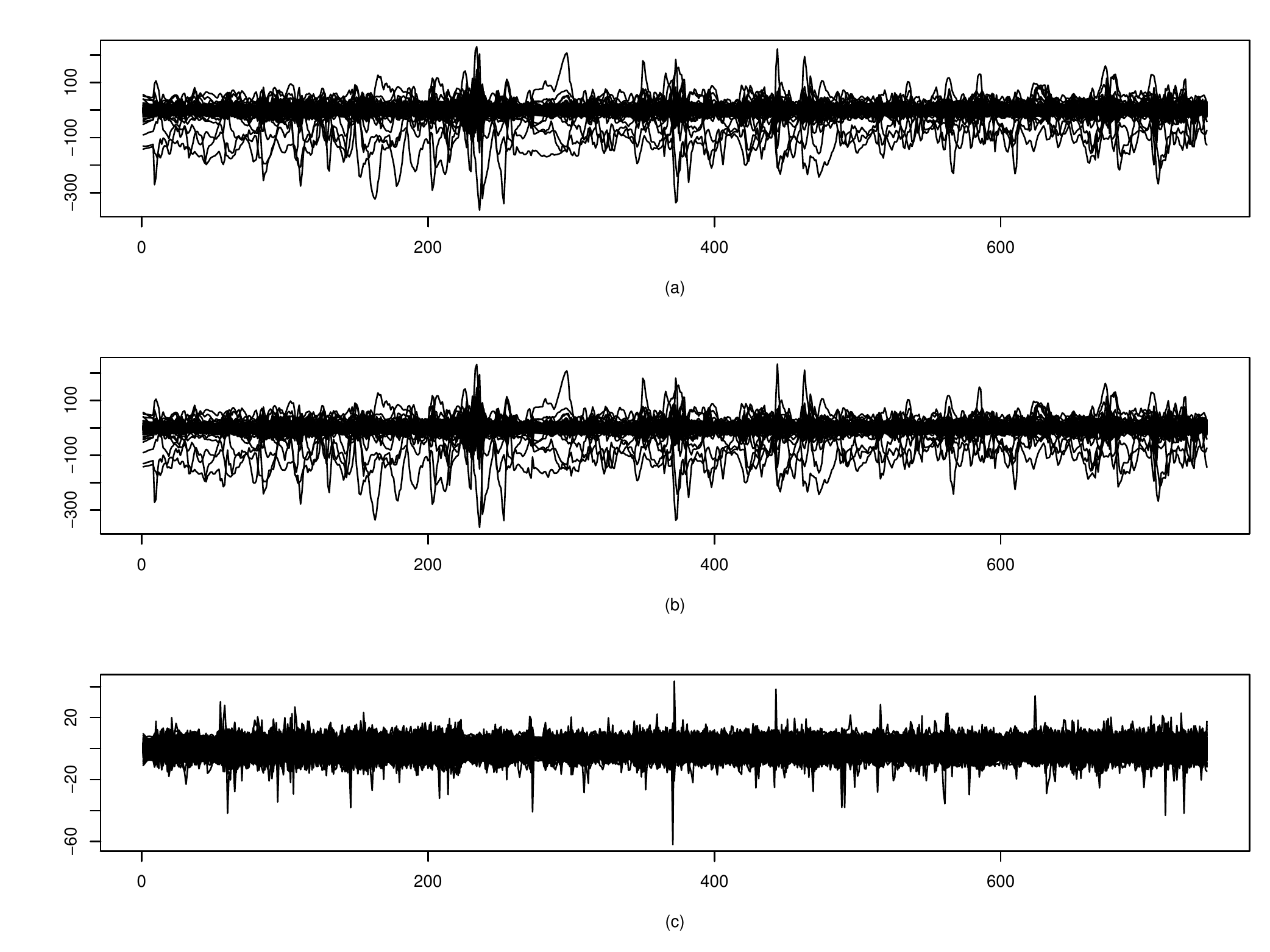}}
\caption{(a) Time plots of the recovered 505-dimensional stationary process $\wh\bx_{2t}$; (b) Time plots of the recovered 256 stationary common factors $\wh\bz_{2t}$; (c) Time plots of the 249 extracted white noise series.}\label{fig8}
\end{center}
\end{figure}

Next, we applied the white noise testing method of Section 2.3 
with $k_0 = j_0 = 2$ for Equations (\ref{wy}) and (\ref{M2}) and found 
that $\wh r_2=256$, i.e., there are 256 stationary common factors and the remaining 
249 components are white noises.  
If we set the parameters $k_0=j_0=1$ and $k_0=j_0=3$,  we obtain that $\wh r_2=292$ and $281$, respectively.  But using these factors produces similar results below and, for simplicity, 
we only report the analysis with $k_0=j_0=2$. To obtain the extracted stationary factors, by the projected PCA in \cite{gaotsay2019b}, we first examine the eigenvalues of the sample covariance matrix $\wh\bS$ defined in (\ref{Spca}).
From Figure~\ref{fig7}(a), we see that there is an eigenvalue of the covariance matrix of the white noise components that is much larger than the others. Therefore, we choose $K=1$ and the recovered stationary factors and the white noise components are shown in Figure~\ref{fig8}(b) and (c), respectively. From Figure~\ref{fig8}, we see that the 256 stationary common factors in part (b) capture most of the nontrivial dynamic dependencies of the stationary components $\wh\bx_{2t}$ shown in part (a), and the remaining 249 white noise series capture little dynamic information of $\wh\bx_{2t}$. 
Consequently, for the 508-dimensional time series of PM$_{2.5}$ measurements, our proposed method  recovers 3 unit-root series of common trends, 256 stationary common factors with 
non-trivial dynamic dependence, and 
249 white noise series which capture part of the contemporaneous variability of the data.

\begin{figure}
\begin{center}
{\includegraphics[width=0.65\textwidth]{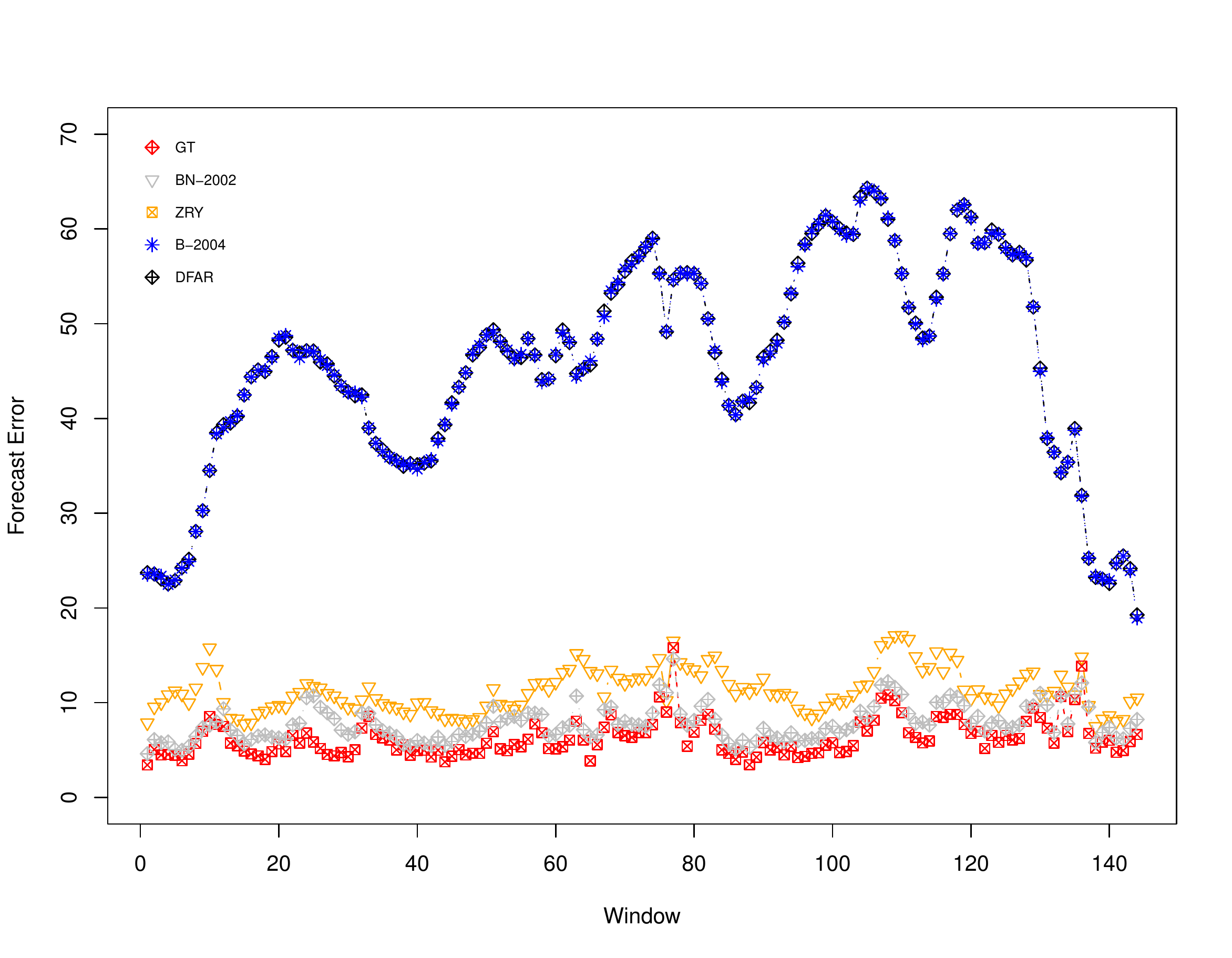}}
\caption{Time plots of the 1-step ahead pointwise forecast errors using various models of Example 3. GT denotes the proposed method, ‘BN-2002’ denotes the stationary approach of  \cite{BaiNg_Econometrica_2002},  B-2004 is the nonstationary method of \cite{bai2004}, ZRY denotes the approach in \cite{zhang-etal2019}, and DFAR is the scalar AR approach to the differenced PM$_{2.5}$ data. }\label{fig9}
\end{center}
\end{figure}

Next we examine  and compare the forecasting performance of the extracted factors via 
the proposed method with different methods available in the literature. 
We estimate the models using the data in the time span $[1,\tau]$ with $\tau=600,...,744-h$ for the $h$-step ahead forecasts, i.e., we use the last 6 days of March 2017 for out-of-sample forecasting. 
First, we applied the stationary factor approach of \cite{BaiNg_Econometrica_2002} and the nonstationary one in \cite{bai2004}, and found that the numbers of common factors are 20 and 1, respectively. We also set the number of nonstationary factors in \cite{zhang-etal2019} (denoted by ZRY) to $\wh{r}_1=3$. 
Second, following the discussion of Section 2.2, we fit a  VARIMA(1,1,0) model to the nonstationary factors $\wh\bx_{1t}$ and scalar AR(1) models to  the stationary common factors $\wh\bz_{2t}$, and denote this approach by GT. Since the dimension of the extracted stationary common factor 
is high, we employ scalar AR models to provide a quick and simple approximation. We also fit a univariate ARIMA(1,1,0) model to the factor process extracted by \cite{bai2004} and a VAR(1) model to the 20-dimensional factors extracted by \cite{BaiNg_Econometrica_2002}, where the insignificant parameters in the coefficient matrix are removed. We denote these two approaches 
by B-2004 and BN-2002, respectively. As a benchmark model, we also employ scalar AR(1) 
models to the differenced PM$_{2.5}$ series and denote the result by DFAR. 
We compute the $h$-step ahead predictions of $\by_t$ using the predictions of factors and the associated factor loadings. The forecast error is defined as
\begin{equation}\label{fe}
\text{FE}_h=\frac{1}{144-h+1}\sum_{\tau=600}^{744-h}E(\tau,h)\quad\text{with}\quad E(\tau,h)=\frac{1}{\sqrt{p}}\|\wh\by_{\tau+h}-\by_{\tau+h}\|_2,
\end{equation}
where $p=508$. 
Table \ref{Table4.21} reports the 1-step to 4-step ahead forecast errors of Equation (\ref{fe}) 
for the various models considered. The smallest forecast error of each step is shown in boldface. From the table, we see that our proposed method is capable of producing accurate forecasts and the associated forecast errors based on the extracted stationary and nonstationary factors by our method are smaller than the factors extracted by other methods. In addition, we note that the ZRY approach 
and the stationary approach of BN-2002 can also produce relatively small forecast errors, 
but  the single nonstationary factor process recovered by \cite{bai2004} is not able to make accurate predictions. The benchmark approach DFAR produces results similar to that of B-2004, 
which are not satisfactory. For further illustration, the pointwise 1-step ahead forecast errors 
of various methods are shown in Figure~\ref{fig9}. From the 
plot, we see that the proposed approach outperforms the other four methods in most time periods 
of the forecasting subsample.

\begin{table}[h]
 \caption{The 1-step to 4-step ahead out-of-sample forecast errors. GT denotes the proposed method, ‘BN-2002’ denotes the stationary approach of \cite{BaiNg_Econometrica_2002},  B-2004 is the nonstationary method of \cite{bai2004}, ZRY denotes the approach of \cite{zhang-etal2019}, and DFAR is the 
 scalar AR approach to the differenced PM$_{2.5}$ data.
 Boldface numbers denote the smallest one for a given model.}
          \label{Table4.21}
\begin{center}
 \setlength{\abovecaptionskip}{0pt}
\setlength{\belowcaptionskip}{3pt}

\begin{tabular}{c|cccccc}
\hline
&&\multicolumn{5}{c}{Methods}\\
\cline{3-7}
Step&&GT&ZRY&B-2004&BN-2002&DFAR\\
\hline
1 &  & {\bf 6.25} & 11.29 & 45.77&7.79&45.80 \\
2& &{\bf 8.58}&12.03&45.93&9.68&45.95\\
3& &{\bf 10.18}&12.82&46.08&11.12&50.33\\
4& &{\bf 11.50}&13.67&46.25&12.26&46.27\\
\hline
\end{tabular}
          \end{center}
\end{table}

We then compare the forecast ability of extracted factors by different approaches and the benchmark method by adopting the asymptotic  test in \cite{diebold-1995} to test for equal predictive ability without a rigorous argument.  The null hypothesis of interest is that the approaches considered have equal predictive ability, and the alternative is that our proposed method performs better than the others in predicting the PM$_{2.5}$ series. 
Table~\ref{Table43} reports the testing results, where we use the method in \cite{andrews1991} to calculate the long-run covariance matrix (L-COV). In the table, (GT)--(ZRY), (GT)--(BN-2002), (GT)--(B-2004), and (GT)--(DFAR) denote the comparisons between our approach and that in ZRY, BN-2002, B-2004, and DFAR, respectively, and the test statistic is the difference between the forecast errors of the two methods involved, as illustrated in \cite{diebold-1995}. From Table~\ref{Table43}, we see that all $p$-values of the test statistics are 
small and less than 0.05, indicating that the factors 
extracted by our method have better forecast ability than those extracted by 
other four methods  in predicting all components of the PM$_{2.5}$ measurements. The results in Tables~\ref{Table4.21} and \ref{Table43} are understandable since the factors extracted by our method capture most of the dynamic dependence of the data by eliminating the white noise effect. These extracted common factors are helpful in predictions.
\begin{table}
\caption{Testing for equal predictive ability of different methods, using the asymptotic test of \cite{diebold-1995}. GT denotes the proposed method, ‘BN-2002’ denotes the stationary approach in \cite{BaiNg_Econometrica_2002},  B-2004 is the nonstationary method in \cite{bai2004}, ZRY denotes the approach in \cite{zhang-etal2019}, and DFAR is the componentwise AR approach to the differenced data.} 
          \label{Table43}
\begin{center}
 \setlength{\abovecaptionskip}{0pt}
\setlength{\belowcaptionskip}{3pt}

\begin{tabular}{c|cccccc}
\hline
Step&&(GT)--(ZRY)&(GT)--(BN-2002)&&(GT)--(B-2004)&(GT)--(DFAR)\\
\hline
1 & L-COV&0.161  & 0.015 & &279.5&  266.1 \\
   & $p$-value& $\approx 0$& $\approx 0$& &$0.01$  &$0.01$ \\ 
   \hline 
2&L-COV&0.154&0.055&&61.48&61.64\\
&$p$-value&$\approx 0$&$\approx 0$&&$\approx 0$&$\approx 0$\\
\hline
3&L-COV&0.263&0.086&&71.14&71.17\\
&$p$-value&$\approx 0$&$\approx 0$&&$\approx 0$&$\approx 0$\\
\hline 
4&L-COV&0.273&0.131&&70.20&70.23\\
&$p$-value&$\approx 0$&$0.017$&&$\approx 0$&$\approx 0$\\
\hline 
\end{tabular}
          \end{center}
\end{table}

To further compare the forecast ability of the nonstationary factors  extracted by the proposed method and the one in \cite{bai2004}, we employ the factor-augmented error correction models (FECM) in \cite{banerjee-etal2014}. As the factors in \cite{banerjee-etal2014} are extracted by the method of \cite{bai2004}, we denote the 
approach of FECM of \cite{banerjee-etal2014} by Bai-BMM, and our method by GT-BMM. Since the method in \cite{bai2004} identifies 1 factor and the proposed method estimates 3, we include 1 to 3 factors in the FECM defined in Equation (8) of \cite{banerjee-etal2014} with $q=1$. That is, we include the lag-1 of the differenced augmented vectors in the error correction model and denote it 
by FECM(1). We study the forecast of the PM$_{2.5}$ in Daliao district of Kaohsiung, which is one of the most polluted  areas in Southern Taiwan. Table~\ref{Table4.22} presents the 1-step to 4-step ahead root mean squared forecast errors (RMSFE) defined as
\begin{equation}\label{rmfe}
\text{RMSFE}_h=\left(\frac{1}{144-h+1}\sum_{\tau=600}^{744-h}(\wh y_{i,\tau+h}-y_{i,\tau+h})^2\right)^{1/2},
\end{equation}
where $y_{i,t}$ is the observed PM$_{2.5}$ data at Daliao district. From Table~\ref{Table4.22}, we see that our nonstationary factors perform as good as the ones in \cite{bai2004} using 
the FECM of \cite{banerjee-etal2014}, and our factors perform slightly better when we adopt 
all 3 detected unit-root common factors. 

In conclusion, for the PM$_{2.5}$ data in Taiwan, the forecasting performance of the extracted factors seems to favor the proposed approach when we adopt the factor-based modeling and make predictions using the associated loadings. The nonstationary factors extracted by our method also fare well in comparison with the one by \cite{bai2004} when using the FECM approach in \cite{banerjee-etal2014}. Finally, we emphasize that the proposed model is different from the ones in \cite{bai2004} and \cite{StockWatson_2002a} because it explores a different aspect of the data. The proposed method is intended as another option for modeling high-dimensional unit-root time series.


\begin{table}[h]
 \caption{The 1-step to 4-step ahead root mean squared forecast errors. GT-BMM denotes the FECM using the factors  of proposed method, Bai-BMM is the FECM approach using the factors in \cite{bai2004}.}
          \label{Table4.22}
\begin{center}
 \setlength{\abovecaptionskip}{0pt}
\setlength{\belowcaptionskip}{3pt}\footnotesize
\begin{tabular}{ccccccccccc}
\hline
&&&\multicolumn{2}{c}{1 factor}&&\multicolumn{2}{c}{2 factors}&&\multicolumn{2}{c}{3 factors}\\
\cline{1-2}\cline{4-5}\cline{7-8}\cline{10-11}
Step&Model&&GT-BMM&Bai-BMM&&GT-BMM&Bai-BMM&&GT-BMM&Bai-BMM\\
\cline{1-2}\cline{4-5}\cline{7-8}\cline{10-11}
1&FECM(1)&&9.08&9.07&&14.45&14.40&&14.28&15.48\\
2&FECM(1)&&13.32&13.26&&18.32&18.33&&18.97&20.58\\
3&FECM(1)&&15.79&15.70&&19.37&19.36&&20.30&21.20\\
4&FECM(1)&&17.22&17.15&&19.98&19.95&&20.75&21.33\\
\hline
\end{tabular}
          \end{center}
\end{table}


\section{Concluding Remarks} 
This paper proposed a new approach to analyze high-dimensional unit-root time series from a factor modeling perspective. The proposed approach marks an extension of the work of \cite{zhang-etal2019}, \cite{gaotsay2019a,gaotsay2019b}, and 
\cite{penaponcela2006}, and is in line with the frameworks of \cite{TiaoTsay_1989} 
and \cite{BoxTiao_1977}. 
Our approach uses modified methods of \cite{zhang-etal2019} to separate
the unit-root series from the stationary components via an eigenanalysis of certain 
nonnegative definite matrix, which combines variance and lagged autocovariance matrices together. 
To further reduce the dimension of the stationary components, we extend the 
method of \cite{gaotsay2019b} with improvements in high-dimensional white noise 
testing. The proposed approach is easy to implement for high-dimensional time series 
and  empirical results show that 
it  can effectively extract the unit root series and stationary common factors from complex data. 
In addition, the extracted common trends and stationary common factors are useful in 
out of sample predictions. Thus, the proposed approach expands the toolkits for 
analyzing high-dimensional unit-root time series. 

\section*{Appendix: Proofs}
\setcounter{equation}{0}
\renewcommand\theequation{A.\arabic{equation}}
Let $\wh\bSigma_1(k)$ and $\wh\bSigma_2(k)$ be the lag-$k$ sample autocovariances of $\wh\bx_{1t}$ and $\wh\bx_{2t}$, respectively. Define
$\bSigma_2(k)=\Cov(\bx_{2,t+k},\bx_{2t})$, $\bSigma_x(k)=\diag(\wh\bSigma_1(k),\bSigma_2(k))$, and
$\bM_x=\sum_{k=0}^{k_0}\bSigma_x(k)\bSigma_x(k)'=:\diag(\bD_1^x,\bD_2^x).$
There exists a $p\times p$ orthogonal matrix $\bGamma_x$ such that 
$\bM_x\bGamma_x=\bGamma_x\bLambda_x$,
where $\bLambda_x$ is a diagonal matrix of eigenvalues (in decreasing order) of $\bM_x$. Since $\bx_{1t}$ is unit-root nonstationary and $\bx_{2t}$ is stationary, $\bD_1^x$ and $\bD_2^x$ do not share the same singular values. Therefore, $\bGamma_x$ must be block-diagonal. Define
$\bM_y=\bA\bM_x\bA'=\bA\bGamma_x\bLambda_x\bGamma_x'\bA',$
which implies that the columns of $\bA\bGamma_x$ are the orthogonal eigenvectors of $\bM_y$. Since $\bGamma_x$ is block-diagonal, it follows that $\mathcal{M}(\bA_1)$ and $\mathcal{M}(\bA_2)$ are the same as the spaces spanned by the eigenvectors corresponding to the largest $r_1$ and the smallest $(p-r_1)$ eigenvalues of $\bM_y$, respectively. Thus, we only need to show that the space spanned by the eigenvectors of $\bM_y$ can be approximated by that of $\wh\bM_1$. This question is usually solved by the perturbation theory. In particular, let
\begin{equation}
\wh\bM_1=\bM_y+\Delta\bM_y,\,\, \Delta\bM_y=\wh\bM_1-\bM_y.
\end{equation}
We adopt the perturbation results of  \cite{dopico2000} in our proofs below, and use $C$ or $c$ as a generic constant whose value may change at different places.

{\noindent\bf Proof of Theorem 1}. By Assumption 2 and (A.2) in the proof of Theorem 1 of \cite{gaotsay2019b}, we can show that
\[\max_{0\leq k\leq k_0}\|\frac{1}{n}(\bx_{2t}-\bar{\bx}_2)(\bx_{2,t-k}-\bar{\bx}_2)'-\Cov(\bx_{2t},\bx_{2,t-k})\|_2=O_p(n^{-1/2}),\]
which implies that Condition 1(ii)  of \cite{zhang-etal2019} holds. Under Assumptions 1-2, by a similar argument as the proof of Theorem 3.1 there, (\ref{Ai}) holds. 
In addition, noting that
\begin{equation}\label{a2y}
\wh\bA_2'\by_t=(\wh\bA_2-\bA_2)'\by_t+\bx_{2t},
\end{equation}
we have 
\begin{align}\label{s2t}
\wt\bSigma_2(j)=&(\wh\bA_2-\bA_2)'\wh\bSigma_y(j)(\wh\bA_2-\bA_2)+(\wh\bA_2-\bA_2)'\wh\bSigma_{y2}(j)+\wh\bSigma_{2y}(j)(\wh\bA_2-\bA_2)+\wh\bSigma_2(j)\notag\\
=:&\bJ_1+\bJ_2+\bJ_3+\bJ_4,
\end{align}
where $\wh\bSigma_{y2}(j)$ and $\wh\bSigma_{2y}(j)$ are the sample covariances of $\by_{t+j}$ and $\bx_{2t}$, and $\bx_{2,t+j}$ and $\by_t$, respectively. Since
\begin{equation}\label{sy:t}
\wh\bSigma_y(j)=\bA\left[\begin{array}{cc}
\wh\bSigma_1(j)&\wh\bSigma_{12}(j)\\
\wh\bSigma_{21}(j)&\wh\bSigma_2(j)
\end{array}\right]\bA',
\end{equation}
where $\wh\bSigma_{12}(j)$ and $\wh\bSigma_{21}(j)$ are the sample covariances of $\bx_{1,t+j}$ and $\bx_{2t}$, and $\bx_{2,t+j}$ and $\bx_{1t}$, respectively, 
by Theorem 1 of \cite{penaponcela2006}, for $0\leq j\leq k_0$,
\begin{equation}\label{sig1h}
\|\wh\bSigma_1(j)\|_2=O_p(n).
\end{equation}
By the proof of Lemma 4 in \cite{zhang-etal2019}, we have
\begin{equation}\label{sig12}
\|\wh\bSigma_{12}(j)\|_2=O_p(n^{1/2})\quad\text{and}\quad \|\wh\bSigma_{21}(j)\|_2=O_p(n^{1/2}).
\end{equation}
Under Assumptions 2 and 3, by the proof of Theorem 1 in \cite{gaotsay2019b},
\begin{equation}\label{sig2:df}
\|\wh\bSigma_2(j)-\bSigma_2(j)\|_2=O_p(n^{-1/2}).
\end{equation}
Therefore, by (\ref{s2t})--(\ref{sig2:df}),
\begin{align}\label{sig2t}
\|\wt\bSigma_2(j)-\bSigma_2(j)\|_2\leq& \|\bJ_1\|_2+\|\bJ_2\|_2+\|\bJ_3\|_2+\|\bJ_4-\bSigma_2(j)\|_2\notag\\
\leq &\|\wh\bA_2-\bA_2\|_2^2\|\wh\bSigma_y(j)\|_2+\|\wh\bA_2-\bA_2\|_2\|\wh\bSigma_{y2}(j)\|_2+\|\wh\bA_2-\bA_2\|_2\|\wh\bSigma_{2y}(j)\|_2\notag\\
&+\|\wh\bSigma_2(j)-\bSigma_2(j)\|_2\notag\\
\leq &O_p(n^{-1}+n^{-1/2}+n^{-1/2}+n^{-1/2})=O_p(n^{-1/2}).
\end{align}
Then, by  a similar argument as the proof of Theorem 3 in \cite{gaotsay2019a},
\begin{equation}\label{u1}
\|\wh\bU_1-\bU_1\|_2\leq C\|\wt\bSigma_2(j)-\bSigma_2(j)\|_2=O_p(n^{-1/2}),
\end{equation}
and hence,
\begin{equation}\label{a2u1}
\|\wh\bA_2\wh\bU_1-\bA_2\bU_1\|_2=O_p(\|\wh\bA_2-\bA_2\|_2+\|\wh\bU_2-\bU_2\|_2)=O_p(n^{-1/2}).
\end{equation}
The proof for the second part of (\ref{A2U1}) is similar. Thus, (\ref{A2U1}) holds.\\
By (A.10) in the supplement of \cite{gaotsay2019b}, (\ref{sig2t}) and (\ref{A2U1}),
\begin{equation}\label{shat:con}
\|\wh\bS-\bS\|_2\leq \|\wt\bSigma_2-\bSigma_2\|_2^2+4\|\bSigma_2\|_2\|\wt\bSigma_2-\bSigma_2\|_2+3\|\bSigma_2\|_2^2\|\wh\bA_2\wh\bV_1-\bA_2\bV_1\|_2=O_p(n^{-1/2}).
\end{equation}
By a similar argument as the proof of (\ref{u1}) above,
\begin{equation}\label{v2hat}
\|\wh\bV_2-\bV_2\|_2\leq C\frac{\|\wh\bS-\bS\|_2}{\lambda_K(\bS)}=O_p(n^{-1/2}).
\end{equation}
which implies (\ref{A2V2}). Finally, by (\ref{a2y}), we have
\begin{equation}\label{x1:ext}
\wh\bA_1\wh\bA_1'\by_t=\wh\bA_1(\wh\bA_1-\bA_1)'\by_t+(\wh\bA_1-\bA_1)\bx_{1t}+\bA_1\bx_{1t}.
\end{equation}
By Lemma 1 of \cite{zhang-etal2019}, $\|\by_t\|_2=O_p(n^{1/2})$ and $\|\bx_{1t}\|_2=O_p(n^{1/2})$, thus,
\[\|\wh\bA_1\wh\bA_1'\by_t-\bA_1\bx_{1t}\|_2\leq\|\wh\bA_1-\bA_1\|_2\|\by_t\|_2+\|\wh\bA_1-\bA_1\|_2\|\bx_{1t}\|_2= O_p(n^{-1/2}).\]
By a similar argument and the proof of Theorem 1 of \cite{gaotsay2019b}, the second result of (\ref{ext:ft}) also holds. This completes the proof.
$\Box$

{\noindent\bf Proof of Theorem 2.} To show $P(\wh r_1=r_1)\rightarrow 1$, it suffices to show that, for any  column $\wh\ba_2$ of $\wh\bA_2$, $\max_{1\leq t\leq n}\|\wh\ba_2'\bA_1\bx_{1t}\|_2=o_p(1)$, because
$\wh\ba_2'\bA_1\bx_{1t}=(\wh\ba_2-\ba_2)'\bA_1\bx_{1t},$
and $\|\wh\ba_1-\ba_2\|_2\leq\|\wh\bA_2-\bA_2\|_2=O_p(n^{-1})$ by Theorem 1. Let $S_{it}=\sum_{j=1}^tw_{ij}$ be the partial sum of $w_{it}$. By Assumption 1 and Theorem 1 of \cite{merlevede2011}, there exists $\nu=\gamma_1\gamma_2/(\gamma_1+\gamma_2)$ such that
\begin{align}\label{bern}
P(\max_{1\leq t\leq n}\|\bx_{1t}\|_2>x)\leq CP(\max_{1\leq t\leq n}|S_{it}|>Cx)
\leq &Cn\exp(-Cx^{\gamma_1})+C\exp(-Cx^2/n)\notag\\
&+C\exp\left(-C\frac{x^2}{n}\exp(\frac{Cx^{\nu(1-\nu)}}{(\log x)^\nu})\right).
\end{align}
Thus, we can choose $x=Cn^{1/2}\log(n)$ such that the above probability is of $o_p(1)$. Therefore, $\max_{1\leq t\leq n}\|\bx_{1t}\|_2=O_p(n^{1/2}\log(n))$. It follows that
\[\max_{1\leq t\leq n}\|\wh\ba_2'\bA_1\bx_{1t}\|_2=O_p(n^{-1/2}\log(n))=o_p(1).\]
Furthermore, note that, for any column $\wh\bv_{1i}$ of $\wh\bV_1$ and the corresponding $\bv_{1i}$ of $\bV_1$, 
\begin{align}\label{vay}
\wh\bv_{1i}'\wh\bA_2'\by_t=&\wh\bv_{1i}'(\wh\bA_2-\bA_2)'\by_t+\wh\bv_{1i}'\bU_1\bz_{2t}+(\wh\bv_{1i}-\bv_{1i})'\bU_2\be_t+\bv_{1i}'\bU_2\be_t\notag\\
=:&\balpha_1+\balpha_2+\balpha_3+\balpha_4.
\end{align}
Under Assumption 3, when the dimension $p$ is finite, for any row $\bq_i'$ of $\bQ_1$ defined in (\ref{re:st}), we have
\begin{equation}\label{zt-sub}
P( \max_{1\leq t\leq n} \|\bz_{2t}\|_2>x)\leq CnP(\|\bq_i'\bff_{2t}\|_2>Cx)\leq Cn\exp(-Cx).
\end{equation}
Therefore, we can choose $x=C\log(n)$ such that the above rate is of $o_p(1)$, which implies that $\max_{1\leq t\leq n} \|\bz_{2t}\|_2=O_p(\log(n))$. Similarly, we can show $\max_{1\leq t\leq n}\|\be_t\|_2=O_p(\log(n))$.
Note that $\max_{1\leq i\leq v}\|\wh\bv_{1i}-\bv_{1i}\|_2\leq \|\wh\bV_1-\bV_1\|_2=O_p(n^{-1/2})$, and the arguments in (\ref{bern}) and (\ref{zt-sub}) and that of $\be_t$ also imply that $\max_{1\leq t\leq n}\|\by_t\|_2=O_p(n^{1/2}\log(n))$.
Then,
\begin{align*}
\max_{1\leq i\leq v}\max_{1\leq t\leq n}\|\balpha_1+\balpha_2+\balpha_3\|_2\leq& Cn^{1/2}\log(n)\|\wh\bA_2-\bA_2\|_2+C\log(n)\|\wh\bV_1-\bV_1\|_2\\
=&O_p(n^{-1/2}\log(n))=o_p(1),
\end{align*}
implying that the effects of the estimators $\wh\bv_{1i}$ and $\bA_2$ on the white noise component $\bv_{1i}'\bU_2\bve_t$ in (\ref{vay}) are asymptotically negligible. We can then consistently estimate the number of white noise components asymptotically using white noise tests. This completes the proof. $\Box$\\
Turn to the proofs when the dimension $p$ is large. For $1\leq i\leq r_1$,  $F^i(t)=W^i(t)-\int_0^1 W^i(t)dt$ and $\bF(t)=(F^1(t),...,F^{r_1}(t))'$. Let $\bD_{1n}=\sqrt{n}\bI_{r_1}$. 

\begin{lemma}
(i) Suppose $x_{it}\sim I(1)$ for $1\leq i\leq r_1$, then under Assumption 5,
\begin{equation}\label{x1t}
\frac{x_{i[nt]}-\bar{x}_i}{p^{(1-\delta)/2}\sqrt{n}}\overset{J_1}{\Longrightarrow}F^i(t),\,\, \text{for}\,\, 1\leq i \leq r_1,\,\,\text{on}\,\, D[0,1];
\end{equation}
(ii) Under the conditions in (i),
\begin{equation}\label{sig:con}
p^{-(1-\delta)}\bD_{1n}^{-1}\wh\bSigma_1(0)\bD_{1n}^{-1}\longrightarrow_d\int_0^1\bF(t)\bF(t)'dt.
\end{equation}
\end{lemma}

{\bf Proof.} For any $I(1)$ process, we can write $\nabla x_{it}=w_{it}$ and 
\begin{equation}\label{xit:r}
x_{it}=\sum_{j=1}^t w_{ij}.
\end{equation}
By Assumption 5 and the continuous mapping theorem, it follows that
\begin{equation}\label{xit:conv}
\frac{x_{i[nt]}-\bar{x_i}}{p^{(1-\delta)/2}\sqrt{n}}\overset{J_1}{\Longrightarrow} W^i(t)-\int_0^1 W^i(t)dt,\,\, \text{for}\,\, 1\leq i \leq r_1,
\end{equation}
which implies (\ref{x1t}). (\ref{sig:con}) follows immediately from (i) by the continuous mapping theorem. This completes the proof of Lemma 1. $\Box$

Let $\wh\bSigma_x(k)=\bA'\wh\bSigma_y(k)\bA$, $\bGamma_x(k)=\diag(\wh\bSigma_1(0),\Cov(\bx_{2,t+k},\bx_{2t}))$, and $\bD_n=\diag(\bD_{1n},\bI_{p-{r_1}})$. Define $\kappa=\max(r_2,K)$, $\bDelta_p=\diag(p^{(1-\delta)/2}\bI_\kappa,\bI_{p-r_1-\kappa})$ and $\bTheta_p=\diag(p^{(1-\delta)/2}\bI_{r_1},\bDelta_p)$

\begin{lemma}\label{lm2}
(i) Under Assumptions 2--6, we have
\begin{equation}\label{x1:appr}
\max_{0\leq k\leq k_0}p^{-(1-\delta)}\|\bD_{1n}^{-1}(\wh\bSigma_1(k)-\wh\bSigma_1(0))\bD_{1n}^{-1}\|_2=O_p(n^{-1/2}).\\
\end{equation}
(ii) Under Assumptions 2--6, we have 
\begin{equation}\label{s2:con1}
\max_{0\leq k\leq k_0}\|\bDelta_p^{-1}(\wh\bSigma_2(k)-\bSigma_2(k))\bDelta_p^{-1}\|_2=O_p(p^{\delta/2}r_2^{1/2}n^{-1/2}),
\end{equation}
and there exists a constant $c>0$ such that $\lambda_{\min}(\bDelta_p^{-1}\bSigma_2(0)\bDelta_p^{-1})\geq c>0.$
\end{lemma}

{\bf Proof.} We prove (\ref{x1:appr}) first. Note that $1\leq i,h\leq r_1$,
\beqn \label{decm}&&\sum_{t=1}^{n-j}(x_{i,t+j}-\bar{x}_i)(x_{ht}- \bar{x}_h)-\sum_{t=1}^{n}(x_{it}-\bar{x}_i)(x_{ht}- \bar{x}_h)\nn\\
&=&
-\sum_{t=1}^{j}(x_{it}-\bar{x}_i)(x_{ht}- \bar{x}_h)-\sum_{t=1}^{n-j}(x_{i,t+j}-\bar{x}_i)
(x_{h,t+j} - \bar{x}_{h})\nn\\
&=:&\delta_{n1}(j, i, h)+\delta_{n2}(j, i, h).\nn\eeqn
It follows from  Lemma 1(i) that
\beqn \label{dt1}\sup_{0\leq j\leq k_0}{|\delta_{n1}(j, i, h)|\over n^{2}p^{1-\delta}}&\leq& {k_0\over n}\left(\sup_{1\leq t\leq n}{|x_{it}-\bar{x}_i|\over \sqrt{n}p^{(1-\delta)/2}}\right)\left(\sup_{1\leq t\leq n}{|x_{ht}-\bar{x}_h|\over \sqrt{n}p^{(1-\delta)/2}}\right)\\
&=&O_p(1/n).\nn\eeqn

As for  $\delta_{n2}(j, i, h),$ we have
 $x_{h,t+j}-x_{ht}=\sum_{i=t+1}^{t+j}w_{i}^{h}$. It follows that
 \beqn \label{dt2}\sup_{0\leq j\leq j_0}{|\delta_{n2}(j, i, h)|\over n^{2}p^{1-\delta}}&\leq&
 \left(\sup_{ t\leq n}{|x_{it}-\bar{x}_i|\over \sqrt{n}p^{(1-\delta)/2}}\right)
 \left({1\over
n^{3/2}p^{(1-\delta)/2}}\sum_{t=1}^{n}\sum_{i=t+1}^{t+j_0}|w_{i}^{h}|\right)\\
&=&O_p(1/n^{1/2}).\nn
\eeqn
(\ref{x1:appr}) follows from (\ref{dt1}) and (\ref{dt2}) since $r_1$ is finite.\\
The result (\ref{s2:con1}) holds because, by Lemma 4 of \cite{gaotsay2019b},
\begin{equation}\label{s2:gt}
\max_{0\leq k\leq k_0}\|\wh\bSigma_2(k)-\bSigma_2(k)\|_2=O_p(p^{1-\delta/2}r_2^{1/2}n^{-1/2}).
\end{equation}
Since the nonzero singular values of $\bU_{22,1}$ and the top $K$ singular values of $\bU_{22,2}$ are of order $p^{(1-\delta)/2}$, we can decompose $\bSigma_{2}(0)$ as
\[\bSigma_2(0)=\bPi\diag(p^{1-\delta}\bI_{r_2+K},\bI_{p-r_1-r_2-K})\bPi',\]
where $\bPi$ is an invertible $(p-r_1)\times (p-r_1)$ matrix and hence the smallest singular value $\sigma_{\min}(\bPi)\geq c>0$. Without loss of generality, we assume $\kappa=r_2$ and let
\begin{equation}\label{dec:pi}
\bPi=\left(\begin{array}{cc}
\bPi_{11}&\bPi_{12}\\
\bPi_{21}&\bPi_{22}
\end{array}\right),
\end{equation}
and hence, $\bDelta_p^{-1}\bSigma_2(0)\bDelta_p^{-1}=\bP\bP'$, where
\begin{equation}\label{s02}
\bP=\left(\begin{array}{cc}
p^{-(1-\delta)/2}\bI_{r_2}&\bf 0\\
\bf 0&\bI_{p-r_1-r_2}
\end{array}\right)\left(\begin{array}{cc}
\bPi_{11}&\bPi_{12}\\
\bPi_{21}&\bPi_{22}
\end{array}\right)\left(\begin{array}{cc}
p^{(1-\delta)/2}\bI_{r_2}&\bf 0\\
\bf 0&\bLambda_K
\end{array}\right),
\end{equation}
and $\bLambda_K=\diag(p^{(1-\delta)/2}\bI_K,\bI_{p-r_1-r_2-K})$.
It is not hard to verify that $\bP$ is non-singular if $\bPi$ is non-singular. Thus, we have $\sigma_{\min}(\bP)\geq c>0$, which implies that $\lambda_{\min}(\bDelta_p^{-1}\bSigma_2(0)\bDelta_p^{-1})\geq c>0.$
This completes the proof. $\Box$\\

\begin{lemma}
Under Assumptions 2--6, we have 
 \begin{equation}\label{sig:dif}
 \max_{0\leq k\leq k_0}\|\bTheta_p^{-1}\bD_n^{-1}(\wh\bSigma_x(k)-\bGamma_x(k))\bD_n^{-1}\bTheta_p^{-1}\|_2=O_p(\max(p^{1/2}n^{-1/2},p^{\delta/2}r_2^{1/2}n^{-1/2})). 
 \end{equation}
\end{lemma}

{\bf Proof.} We first show it for the nonstationary block. By Lemma \ref{lm2},
\begin{equation}\label{non-blk}
p^{-(1-\delta)}\|\bD_{1n}^{-1}(\wh\bSigma_1(k)-\wh\bSigma_1(0))\bD_{1n}^{-1}\|_2=O_p(n^{-1/2}).
\end{equation}
Next, we consider the cross-block. Note that for $0\leq j\leq k_0$
\begin{align}\label{x12}
\sum_{t=1}^{n-j}(\bx_{1,t+j}-\bar{\bx}_1)(\bx_{2t}-\bar{\bx}_2)'=&\sum_{t=1}^{n-j}(\bx_{1,t+j}-\bar{\bx}_1)\{\bU_{22,1}(\bff_{2t}-\bar{\bff}_2)+\bU_{22,2}(\bve_t-\bar{\bve})\}'\notag\\
=&\sum_{t=1}^{n-j}(\bx_{1,t+j}-\bar{\bx}_1)(\bff_{2t}-\bar{\bff}_2)'\bU_{22,1}'\notag\\
&+\sum_{t=1}^{n-j}(\bx_{1,t+j}-\bar{\bx}_1)(\bve_t-\bar{\bve})'\bU_{22,2}'.
\end{align}
For $1\leq i \leq r_1$ and $1\leq h\leq r_2$, let
\begin{align}\label{xf}
w_{ih}:=&\frac{1}{n}\sum_{t=1}^{n-j}(x_{i,t+j}-\bar{x}_i)(f_{2,ht}-\bar{f}_{2,h})
\end{align}
and $\bOmega_1=(w_{ih})_{r_1\times r_2}$. Under Assumptions 2--6, by (\ref{xf}) and Lemma 1,
\begin{align}\label{ejh}
\mathrm{E}\sum_{i=1}^{r_1}\sum_{h=1}^{r_2}(\frac{w_{ih}^2}{n})^2\leq& Cr_2n^{-2}np^{1-\delta}\sum_{t,t'=1}^{n-j}\mathrm{E}[\frac{(x_{i,t+j}-\bar{x}_i)}{\sqrt{n}p^{(1-\delta)/2}}\frac{(x_{i,t'+j}-\bar{x}_i)}{\sqrt{n}p^{(1-\delta)/2}}]\notag\\
&\times \mathrm{E}[(f_{2,ht}-\bar{f}_{2h})(f_{2,ht'}-\bar{f}_{2h})]\notag\\
\leq &Cp^{1-\delta}r_2,
\end{align}
which implies that 
\begin{equation}\label{o1}
\|\bOmega_1\|_2=O_p(p^{(1-\delta)/2}r_2^{1/2}).
\end{equation}
Therefore,
\begin{equation}\label{cro:o1}
p^{-(1-\delta)}\|\bD_{1n}^{-1}\bOmega_1\bU_{22,1}\|_2=O_p(r_2^{1/2}n^{-1/2}).
\end{equation}

By a similar argument as above, we can show that
\begin{equation}\label{xf:e}
p^{-(1-\delta)}\|\bD_{1n}^{-1}\sum_{t=1}^{n-j}(\bx_{1,t+j}-\bar{\bx}_1)(\bve_{t}-\bar{\bve})'\bU_{22,2}'\|_2=O_p(p^{1/2}n^{-1/2}).
\end{equation}
It follows from (\ref{x12}), (\ref{cro:o1}) and (\ref{xf:e}) that
\begin{equation}\label{x12:1}
p^{-(1-\delta)}\|\bD_{1n}^{-1} n^{-1}\sum_{t=1}^{n-j}(\bx_{1,t+j}-\bar{\bx}_1)(\bx_{2t}-\bar{\bx}_2)'\|_2=O_p(p^{1/2}n^{-1/2}).
\end{equation}
By a similar argument, we can show that
\begin{equation}\label{x12:2}
p^{-(1-\delta)}\| n^{-1}\sum_{t=1}^{n-j}(\bx_{2,t+j}-\bar{\bx}_2)(\bx_{1t}-\bar{\bx}_1)'\bD_{1n}^{-1}\|_2=O_p(p^{1/2}n^{-1/2}).
\end{equation}
As for the stationary block, by (\ref{s2:gt}), 
\begin{align}\label{sta}
p^{-(1-\delta)}\|\frac{1}{n}\sum_{t=1}^{n-k}(\bx_{2,t+k}-\bar{\bx}_2)(\bx_{2,t}-\bar{\bx}_2)'-\Cov(\bx_{2,t+k},\bx_{2t})\|_2=&O_p(p^{-(1-\delta)}p^{1-\delta/2}r_2^{1/2}n^{-1/2})\notag\\
=&O_p(p^{\delta/2}r_2^{1/2}n^{-1/2}).
\end{align}
Lemma 3 follows from (\ref{non-blk}), (\ref{x12:1})--(\ref{sta}). This completes the proof. $\Box$\\

{\bf Proof of Theorem 3.} By a similar argument as the proof of Theorem 3.1 in \cite{zhang-etal2019} and Theorem 2.4 of \cite{dopico2000},
\begin{equation}\label{da2}
D(\mathcal{M}(\wh\bA_2),\mathcal{M}(\bA_2))\leq (\|\bM_x^{-1/2}\wh\bM_x^{1/2}\|_F+\|\bM_x^{1/2}\wh\bM_x^{-1/2}\|_F)/\eta,
\end{equation}
where $\eta=\min_{\lambda\in\lambda(\bD_1^x),\mu\in\lambda(\wt\bD_2^x)}|\lambda-\mu|/\sqrt{\lambda\mu}$ and $\lambda(\wt\bD_2^x)$ consists of the $p-r_1$ smallest eigenvalues of $\wh\bM_x:=\bA'\wh\bM_1\bA$. By Lemmas 2 and 3, $\eta\geq Cn$. Note that
\[\|(\bM_x)^{-1/2}(\wh\bM_x)^{1/2}\|_F\leq \sum_{j=0}^{k_0}\|(\bSigma_x(0))^{-1}\{\wh\bSigma_x(j)\wh\bSigma_x(j)'\}^{1/2}\|_F.\]
By Lemmas 2 and 3, the solutions ($\lambda_i$, $1\leq i\leq p$) of 
\[|\bTheta_p^{-1}{\bD_n^{-1}}(\wh\bSigma_x(j)\wh\bSigma_x(j))^{1/2}\bD_n^{-1}\bTheta_p^{-1}-\lambda\bTheta_p^{-1}\bD_n^{-1}\bSigma_x(0)\bD_n^{-1}\bTheta_p^{-1}|=0\]
are all bounded in probability. Thus,
\[\|\bM_x^{-1/2}\wh\bM_x^{1/2}\|_F\leq O_p((\sum_{i=1}^p\lambda_i^2)^{1/2})=O_p(p^{1/2}).\]
Similarly, we can show that
$\|\bM_x^{1/2}\wh\bM_x^{-1/2}\|_F=O_p(p^{1/2}).$
Therefore, by Theorem 2.4 of Dopico et al. (2000),
$D(\mathcal{M}(\wh\bA_2),\mathcal{M}(\bA_2))=O_p(p^{1/2}n^{-1}).$ 
The result for $D(\mathcal{M}(\wh\bA_1),\mathcal{M}(\bA_1))$ can be shown similarly  by switching the positions of $\bA_1$ and $\bA_2$. This proves (\ref{Ai:p}).\\

Turn to the estimation of $\bU_1$. First, if $p=o(n^{1/(1+\delta)})$, it is not hard to show that
\[\|\wh\bSigma_1(j)\|_2=O_p(p^{1-\delta}n),\|\wh\bSigma_{12}(j)\|_2=O_p(p^{1-\delta/2}n^{1/2}),\|\wh\bSigma_{21}(j)\|_2=O_p(p^{1-\delta/2}n^{1/2}),\]
\[\|\wh\bSigma_2(j)\|_2=O_p(p^{1-\delta}),\|\wh\bSigma_{y2}(j)\|_2=O_p(p^{1-\delta/2}n^{1/2}),\,\,\text{and}\,\,\|\wh\bSigma_{2y}(j)\|_2=O_p(p^{1-\delta/2}n^{1/2}).\]
By (\ref{sy:t}) and using the above convergence rates, we have 
$\|\wh\bSigma_y(j)\|_2=O_p(p^{1-\delta}n).$
Then, if $r_2=o(\min(p^\delta,p^{1-\delta}))$,  by (\ref{sig2t}) and (\ref{s2:gt}), 
\[\|\wt\bSigma_2(j)-\bSigma_2(j)\|_2=O_p((p^{1/2}n^{-1})^2p^{1-\delta}n+(p^{1/2}n^{-1})(p^{1-\delta/2}n^{1/2})+p^{1-\delta/2}r_2^{1/2}n^{-1/2})=O_p(p^{3/2-\delta/2}n^{-1/2}).\]
By the proof in Lemma 5 of \cite{gaotsay2019b},
\[\|\wh\bM_2-\bM_2\|_2=O_p(\|\wt\bSigma_2(j)-\bSigma_2(j)\|_2^2+\|\wt\bSigma_2(j)-\bSigma_2(j)\|_2\|\bSigma_2(j)\|_2)=O_p(p^{5/2-3\delta/2}n^{-1/2}).\]
Note that $\lambda_{r_2}(\bM_2)\geq Cp^{2(1-\delta)}$. By a similar argument as (\ref{u1}),
\[\|\wh\bU_1-\bU_1\|_2\leq C\frac{\|\wh\bM_2-\bM_2\|_2}{\lambda_{r_2}(\bM_2)}=O_p(p^{(1+\delta)/2}n^{-1/2}).\]
(\ref{A2U1:p}) follows from (\ref{Ai:p}) and the above result.\\
By (\ref{shat:con}) and the rates above, if $p=o(n^{1/(1+\delta)})$, we can show that
\begin{align}\label{shat:p}
\|\wh\bS-\bS\|_2=&O_p\{(p^{3/2-\delta/2}n^{-1/2})^2+p^{1-\delta}(p^{3/2-\delta/2}n^{-1/2})+p^{2(1-\delta)}p^{(1+\delta)/2}n^{-1/2}\}\notag\\
=&O_p(p^{5/2-3\delta/2}n^{-1/2}).
\end{align}
Note that $\lambda_K(\bS)\geq p^{2(1-\delta)}$, hence,
\begin{equation}\label{V2st}
\|\wh\bV_2^*-\bV_2^*\|_2\leq C\frac{\|\wh\bS-\bS\|_2}{\lambda_K(\bS)}=O_p(p^{(1+\delta)/2}n^{-1/2}).
\end{equation}
(\ref{A2V2:p}) follows from (\ref{V2st}) and (\ref{Ai:p}).\\
By (\ref{x1:ext}),
\begin{align}\label{x1:ex:p}
\wh\bA_1\wh\bA_1'\by_t-\bA_1\bx_{1t}=&\wh\bA_1(\wh\bA_1-\bA_1)'\bA_1\bx_{1t}+\wh\bA_1(\wh\bA_1-\bA_1)'\bA_2\bU_{22,1}\bff_{2t}\notag\\
&+\wh\bA_1(\wh\bA_1-\bA_1)'\bA_2\bU_{22,2}\bve_t+(\wh\bA_1-\bA_1)\bx_{1t}\notag\\
=:&R_1+R_2+R_3+R_4.
\end{align}
Note that 
\[\|R_1\|_2=O_p(\|\wh\bA_1-\bA_1\|_2\|\bx_{1t}\|_2)=O_p(p^{1/2}n^{-1}p^{(1-\delta)/2}n^{1/2})=O_p(p^{1-\delta/2}n^{-1/2}),\]
\[\|R_2\|_2=O_p(\|\wh\bA_1-\bA_1\|_2\|\bU_{22,1}\|_2\|\bff_{2t}\|_2)=O_p(p^{1/2}n^{-1}p^{(1-\delta)/2}r_2^{1/2})=O_p(p^{1-\delta/2}r_2^{1/2}n^{-1}),\]
\[\|R_3\|_2=O_p(\|\wh\bA_1-\bA_1\|_2\|\bU_{22,2}\bve_t\|_2)=O_p(p^{1/2}n^{-1}p^{(1-\delta)/2}p^{1/2})=O_p(p^{3/2-\delta/2}n^{-1}),\]
\[\|R_4\|_2=O_p(\|\wh\bA_1-\bA_1\|_2\|\bx_{1t}\|_2)=O_p(p^{1/2}n^{-1}p^{(1-\delta)/2}n^{1/2})=O_p(p^{1-\delta/2}n^{-1/2}).\]
Thus, if $p=o(n^{1/(1+\delta)})$ and $r_2=o(\min(p^\delta,p^{1-\delta}))$,
\[p^{-1/2}\|\wh\bA_1\wh\bA_1'\by_t-\bA_1\bx_{1t}\|_2=O_p(p^{(1-\delta)/2}n^{-1/2})=o_p(1).\]
By the proof of Theorem 4 in \cite{gaotsay2019b},
\begin{align}\label{Puz}
p^{-1/2}\|\wh\bU_1\wh\bz_{2t}-\bU_1\bz_{2t}\|_2\leq &Cp^{-\delta/2}r_2^{1/2}\|\wh\bU_1-\bU_1\|_2+Cp^{-\delta/2}\|\wh\bV_2^*-\bV_2^*\|_2+p^{-1/2}\notag\\
=&O_p(p^{1/2}r_2^{1/2}n^{-1/2}+p^{-1/2}).
\end{align}
Therefore, if $p=o(n^{1/(1+\delta)})$ and $r_2=o(\min(p^\delta,p^{1-\delta}))$, 
\begin{align}\label{pa2u1}
p^{-1/2}\|\wh\bA_2\wh\bU_1\wh\bz_{2t}-\bA_2\bU_1\bz_{2t}\|_2\leq& p^{-1/2}\|\wh\bU_1\wh\bz_{2t}-\bU_1\bz_{2t}\|_2+p^{-1/2}\|\wh\bA_2-\bA_2\|_2\|\bU_1\bz_{2t}\|_2\notag\\
\leq &O_p(p^{1/2}r_2^{1/2}n^{-1/2}+p^{-1/2}+p^{(1-\delta)/2}r_2^{1/2}n^{-1})\notag\\
=&O_p(p^{1/2}r_2^{1/2}n^{-1/2}+p^{-1/2}).
\end{align}
This completes the proof. $\Box$

{\bf Proof of Theorem 4.} We first show $P(\wh r_1=r_1)\rightarrow 1$ as $n\rightarrow\infty$. For any column $\wh\ba_{2,i}$ of $\wh\bA_2$, note that
\begin{equation}\label{a2iy}
\wh\ba_{2,i}'\by_t=(\wh\ba_{2,i}-\ba_{2,i})'\bA_1\bx_{1t}+(\wh\ba_{2,i}-\ba_{2,i})'\bA_2\bx_{2t}+\ba_{2,i}'\bA_2\bx_{2t}.
\end{equation}
By the proofs of Theorems 2 and 3 above, we have 
\begin{align}\label{aa1}
\max_{1\leq i\leq p-r_1}\max_{1\leq t\leq n} \|(\wh\ba_{2,i}-\ba_{2,i})'\bA_1\bx_{1t}\|_2\leq& \|\wh\bA_2-\bA_2\|_2\max_{1\leq t\leq n}\|\bx_{1t}\|_2\notag\\
=&O_p(p^{1/2}n^{-1}(p^{(1-\delta)/2}n^{1/2}\log(n)))\notag\\
=&O_p(p^{1-\delta/2}n^{-1/2}\log(n)),
\end{align}
\begin{align}\label{aa2}
\max_{1\leq i\leq p-r_1}\max_{1\leq t\leq n} \|(\wh\ba_{2,i}-\ba_{2,i})'\bA_2\bx_{2t}\|_2\leq& \|\wh\bA_2-\bA_2\|_2\max_{1\leq t\leq n}\|\bA_2\bx_{2t}\|_2\notag\\
=& \|\wh\bA_2-\bA_2\|_2\{\max_{1\leq t\leq n}\|\bA_2\bU_{22,1}\bff_{2t}\|_2+\max_{1\leq t\leq n}\|\bA_2\bU_{22,2}\bve_{t}\|_2\}\notag\\
=&O_p\{p^{1/2}n^{-1}(p^{(1-\delta)/2}r_2^{1/2}\log(nr_2)+p^{1/2}\log(np))\}\notag\\
=&O_p(pn^{-1}\log(np)).
\end{align}
Therefore, if $p=o(n^{1/(1+\delta)})$, then $pn^{-1}\log(np)=o(1)$, and
\begin{equation}\label{a2y:asy}
\wh\ba_{2,i}\by_t=\ba_{2,i}'\bA_2\bx_{2t}+o(1).
\end{equation}
Thus, we can consistently estimate $r_1$ by the proposed methods in Section 2.3.\\
We now prove $P(\wh r_2=r_2)\rightarrow 1$. Note that
\begin{align}\label{yt:max}
\max_{1\leq t\leq n}\|\by_t\|_2\leq & \max_{1\leq t\leq n}\|\bA_1\bx_{1t}\|_2+\max_{1\leq t\leq n}\|\bA_2\bU_{22,1}\bff_{2t}\|_2+\max_{1\leq t\leq n}\|\bA_2\bU_{22,2}\bve_{t}\|_2\notag\\
= &O_p(p^{(1-\delta)/2}n^{1/2}\log(n))+O_p(p^{(1-\delta)/2}r_2^{1/2}\log(nr_2))+O_p(p^{1/2}\log(np))\notag\\
=&O_p(p^{(1-\delta)/2}n^{1/2}\log(n)).
\end{align}
Thus, by (\ref{vay}) and the above results, 
\[\max_{1\leq i\leq v}\max_{1\leq t\leq n}\|\balpha_1\|_2=O_p(\|\wh\bA_2-\bA_2\|_2\max_{1\leq t\leq n}\|\by_t\|_2)=O_p(p^{1-\delta/2}n^{-1/2}\log(n)),\]
\[\max_{1\leq i\leq v}\max_{1\leq t\leq n}\|\balpha_2\|_2=O_p(\|\wh\bV_1-\bV_1\|_2\max_{1\leq t\leq n}\|\bU_1\bz_{2t}\|_2)=O_p(pr_2^{1/2}n^{-1/2}\log(nr_2)),\]
\[\max_{1\leq i\leq v}\max_{1\leq t\leq n}\|\balpha_3\|_2=O_p(\|\wh\bV_1-\bV_1\|_2\max_{1\leq t\leq n}\|\bU_{22,2}\bve_{t}\|_2)=O_p(p^{1+\delta/2}n^{-1/2}\log(np)).\]
Therefore, if $p^{1+\delta/2}n^{-1/2}\log(np)=o(1)$,  the effects of the estimators $\wh\bv_{1i}$ and $\bA_2$ on the white noise component $\bv_{1i}'\bU_2\bve_t$ in (\ref{vay}) are asymptotically negligible. Then, we can consistently estimate the number of white noise components as the sample 
size increases using white noise tests. This completes the proof. $\Box$\\



\end{document}